\documentclass[]{interact}

\usepackage{setspace}
\usepackage{graphicx}
\usepackage{amsmath}
\usepackage{bibentry}
\usepackage{import}
\usepackage{multirow}
\usepackage{booktabs}

\usepackage{comment}
\usepackage{xcolor}
\usepackage{xfrac}
\usepackage{amsmath,bm} 
\usepackage{epstopdf}
\usepackage{subfigure}
\usepackage{caption}
\usepackage{lineno}
\usepackage{ulem}
\usepackage{hyperref}

\usepackage{xspace}

\usepackage[most]{tcolorbox}

\newcommand{\citep}[1]{{\cite{#1}}}

\newcommand{\dt}{\dftl{t}}

\newcommand{\be}{\begin{equation}}
\newcommand{\ee}{\end{equation}}
\newcommand{\bes}{\begin{equation*}}
\newcommand{\ees}{\end{equation*}}
\newcommand{\bse}{\begin{subequations}}
\newcommand{\ese}{\end{subequations}}

\tikzset{cross/.style={cross out, draw=black, minimum size=2*(#1-\pgflinewidth), inner sep=0pt, outer sep=0pt},
cross/.default={1pt}}
\newcommand{\tikzcross}[2][mat-2,fill=mat-2]{\tikz[baseline=-0.6ex]\draw[line width=0.5mm] (0,0.1)--(0.2,-0.1);\hspace{-2.1ex}\tikz[baseline=-0.6ex]\draw[line width=0.5mm] (0,-0.1)--(0.2,0.1);}

\newcommand{\re}[1]{$Re=#1$}
\newcommand{\aoa}[1]{$AoA=#1^o$}
\newcommand{\naca}[1]{NACA#1}

\newcommand\etal[1]{#1~\textit{et al.}}

\makeatletter
\newcommand*{\rom}[1]{\expandafter\@slowromancap\romannumeral #1@}
\makeatother

\newcommand{\vect}[1]{\bm{#1}}

\def \nek {Nek5000}

\def\ee{{\hat {\underline e}}}

\def\dt{ \Delta t }

\def\scriptO{{{\it O}\kern -.42em {\it `}\kern + .20em}}
\def\RR{{{\rm l}\kern - .15em {\rm R} }}
\def\PP{{{\rm l}\kern - .15em {\rm P} }}
\def\L2{{{\sf L}^2}}
\def\H1{{{\sf H}^1}}
\def\PN2{{\PP_{N}-\PP_{N-2}}}

\def\complex{{{\rm C} \kern - .53em {\rm l} \kern + .38em}}

\def\a1{{ | \lambda_{\min} |}}

\def\l1{{   \lambda_{\min}  }}

\def\bu0{{\underline {\bf 0}}}

\def\bu{{\bf u}}

\def\Oh{{\hat \Omega}}

\def\u0{{\underline 0}}

\newcommand{\pp}[2]{\frac{\partial #1}{\partial #2} }

\title{DDES Study of Confined and Unconfined NACA Wing Sections \\ Using Spectral Elements}

\author{ 
  \name{Vishal Kumar\textsuperscript{a}, 
        Ananias Tomboulides\textsuperscript{a,b}, 
        Paul Fischer\textsuperscript{a,c,d} and %
        Misun Min\textsuperscript{a}\footnote{CONTACT Misun Min, Email:\email{mmin@mcs.anl.gov}}
       }
  \affil{\textsuperscript{a}Mathematics and Computer Science, Argonne National Laboratory, USA \\ %
         \textsuperscript{b}Mechanical Engineering, Aristotle University of Thessaloniki, Greece \\%
         \textsuperscript{c}Computer Science, University of Illinois Urbana-Champaign, USA \\
         \textsuperscript{d}Mechanical Science \& Engineering, University of Illinois Urbana-Champaign, USA} 
}
\begin{document}

\maketitle

\begin{abstract}

We develop hybrid RANS-LES strategies within the spectral element code \nek{}
based on the $k-\tau$ class of turbulence models. 
We chose airfoil sections at small flight configurations as our target problem
to comprehensively test the solver accuracy and performance.  
We present verification and validation results of an unconfined
NACA0012 wing section in a pure RANS and in a hybrid RANS-LES setup for an angle of
attack ranging from 0 to 90 degrees.  
The RANS results shows good corroboration with existing experimental and numerical
datasets for low incoming flow angles.  A small discrepancy appears at higher
angle in comparison with the experiments, which is in line with our expectations
from a RANS formulation.  
On the other hand, DDES captures both the attached and separated flow dynamics
well when compared with available numerical datasets. 
We demonstrate that for the hybrid turbulence modeling approach a high-order
spectral element discretization converges faster (i.e., with less resolution)
and captures the flow dynamics more accurately than representative low-order
finite-volume and finite-difference approaches. 
We also revise some of the guidelines on sample size requirements for statistics
convergence.
% 
%, wherein we show that the present formulation requires one-third of the 
%recommended simulation time for low-order, fully implicit time advancement
%schemes.  
%
Furthermore, we analyze some of the observed discrepancies of our unconfined 
DDES at higher angles with the experiments by evaluating the side wall ``blocking"
effect. We carry out additional simulations in a confined `numerical wind
tunnel' and assess the observed differences as a function of Reynolds number.

%We observe differences in drag coefficient between the confined and the
%unconfined cases increase as the Reynolds number is increased. We point to 
%the limitation of blocking corrections employed in experiments and propose the 
%numerical study in a confined setting to be a viable strategy to address potential 
%mismatch between numerical and experimental results. 

\end{abstract}


%\linenumbers

\section{Introduction}

%We develop hybrid RANS-LES strategies in the spectral element code \nek{} for 
%application to a NACA wing geometry and present a validation and verification 
%study of our modeling approaches. We focus on testing the accuracy and 
%performance of \nek{} for NACA0012 wing sections. 
%A brief introduction on the topic is provided below.

Resolved large-eddy simulation (LES) of practical flows remains infeasible in
the near future because of the computational requirement of the formulation in
capturing turbulent scales \citep{Piomelli2014,Choi2012}.
The failure of steady and unsteady Reynolds-averaged Navier--Stokes (RANS) 
turbulence closures for separated flow, on the other hand, is well documented
\citep{Spalart2009a}.  Finding a closure strategy with the simplicity and
elegance of RANS closures and the accuracy of resolved LES has long been a
prominent research question in the turbulence community.  Development of hybrid
RANS-LES (HRLES) strategies in the late 1990s was a big leap toward filling
this gap.  The basic idea behind a hybrid approach is to explicitly switch
between the RANS and the LES formulations based on local grid spacing and
turbulent length scales.  Since \cite{Spalart1997a} first
proposed and \cite{Travin2000} later formalized a hybrid
formulation for turbulence closure, termed detached eddy simulation (DES),
persistent strides have been made in its development, particularly for
aeronautical applications.

Several new modifications to the original DES formulation are currently popular
in the computational fluid dynamics (CFD) community.
These address some inherent problems associated with the original DES~\cite{Menter2003}. 
One major issue is a phenomenon termed grid-induced separation. 
It is an artificial flow separation that is caused by refinement of the grid inside 
the wall boundary layer. 
Such a refinement causes the model to switch from RANS to LES without balancing the 
reduction in eddy viscosity by resolved turbulence~\cite{Spalart2006}. 
 The correction, postulated in \cite{Menter2003}, was to use functions 
to ``shield" the RANS model from the DES formulation inside the boundary layer.  
This new model, termed delayed detached eddy simulation (DDES), remains 
impervious to arbitrary grid refinements.  
Although corrections were originally calibrated for the Spalart--Allmaras model, Gritskevich et al.~\cite{Gritskevich2012} recalibrated the blending coefficients to better suit 
the $k-\omega$ class of turbulence models.  
In this work we utilize the $k-\tau$ model of \cite{Tomboulides2018} 
and its shear stress transport (SST) formulation 
for RANS closure.  Using $\tau$ instead of $\omega$ obviates the need for an ad hoc
boundary condition on $\omega$ near solid walls, which is particularly helpful for nodal-based 
methods such as finite-element and spectral element methods~\cite{dfm02}. We also use the modification proposed
in \cite{Gritskevich2012} for the hybrid formulation.

The DES prediction generally outperforms its unsteady/steady RANS counterpart.
Since HRLES methods were originally formulated for massively separated flows, 
most validation and verification studies prefer simulating flow over
four-digit NACA wings in a stalled flow regime.  In fact, NACA0012 has served as the preferred geometry
by various studies \citep{Shur1999,Strelets2001,Im2014,Bidadi2023}.  The reason is 
also in part because of the availability of good-quality wind tunnel datasets.  
In their original DES study, Shur et al.~\cite{Shur1999} conducted simulations of flow over a 2D \naca{0012} wing 
at \re{100,000} for an angle of attack in the range $0--90^o$.  They observed the superiority 
of DES over their unsteady RANS counterparts within the separated flow regime. 
They also observed good agreement with the experiments for aerodynamic forces, 
although they performed calculations at an order of magnitude lower Reynolds number. 
The justification provided was that flow beyond the stall would remain insensitive to the Reynolds number.        
This argument, however, was not supported by recent study 
\cite{Bidadi2023} on the same geometry. The researchers used an improved wall-modeling
variant of DDES (termed IDDES) and found significantly large differences 
$(\sim 10\%)$ with experiment in both lift and drag coefficients.       

Assessing the accuracy of HRLES methods for massively separated aerodynamic flows is
a challenging task.  The primary reason is improper guidelines on domain
size, resolution, and statistics sample requirements.  During the initial surge 
of DES-based studies, researchers customarily used a spanwise domain size of $1c$ with $\sim 200$
convective times of statistics collection, such as the one in \cite{Strelets2001}. 
In a later study, using \naca{0021} at \aoa{60},
Garbaruk et al.~\cite{Garbaruk2009}
established that simulations with spanwise domain sizes $<4c$ showed 
significant variation in results.  These authors also pointed out that smaller-domain 
simulations needed very large statistics collection time
$\sim \mathcal{O}(10^3)$ to get converged results. This recommendation was for
low-order codes using implicit time advancement schemes. These ideas
need to be retested for high-order schemes using semi-implicit/explicit
time advancement. 
Proper validation with wind tunnel experiments is even more challenging. 
Two main hindrances are the freestream turbulence/flow tripping
and blockage effect caused by the side walls of the wind tunnels.   
Although wall corrections have been proposed by the experimental community,
these are primarily empirical relations at best and do not take into consideration the
actual open-air CFD simulations. One possible solution would be for numericists
to perform simulations in wind-tunnel-like domains and compare the data with the
uncorrected data of the experiments. Several attempts have been made in this regard (for
example, see \cite{Melber-Wilkending2007,Hantrais-GervoisJean-FrancoisPiat2012}), 
but the strategy remains practically unused.

Literature involving HRLES implementations in high-order codes is limited.  
Earlier works on DES and its variants were based on ``mixed-order'' discretization schemes 
\cite{Constantinescu2000,Travin2000,Shur1999}.  While the convective term used third- or
fifth-order schemes, other terms of the Navier--Stokes were discretized by using classic
second-order central difference schemes.  Moreover, the majority of these works used implicit 
upwind schemes that were criticized for their excessive dissipation, resulting in 
suboptimal accuracy for the ``LES'' part in HRLES.  
The work by \cite{Strelets2001} sought to fix the issue by having a 
``blended" discretization to achieve upwind difference only in the RANS region 
and central differencing in LES regions away from the body. Im and Zha \cite{Im2014} compared the accuracy of second- and fourth-order schemes with DDES 
and found the latter to be able to capture finer details of the flow structures and 
the aerodynamic forces accurately. Gao and Li~\cite{Gao2017} implemented DDES within a spectral 
difference method and found results to be more accurate than low-order codes on similar 
grid resolutions. 

The current work focuses on the following goals: 
\textit{(1)} implementation of DDES in a spectral element method
(SEM)-based framework. This implementation is based on the recently proposed
k-$\tau$ SST RANS model of \etal{Tomboulides}~\cite{Tomboulides2024}. 
To the best of our knowledge, this is the first
such implementation ever reported for any SEM-based solver. 
\textit{(2)} in-depth verification and validation of the $k-\tau$ SST based DDES
implementation by comparison with other numerical and experimental studies , and
\textit{(3)} assessing the blocking effect of side walls in wind tunnel experiments 
through simulations of numerically confined configurations.  
This paper is organized as follows.
Section~\ref{sec:methodology} discusses the numerical framework used for this
work. Section~\ref{sec:results} presents the results. These include
extensive validation results for k-$\tau$ based RANS and approaches.
In Section~\ref{sec:conclusion} we conclude our study
by outlining the current limitations and necessary future work.

\section{Numerical framework \label{sec:methodology}}

Nek5000 is a spectral element method code that is used for a wide range 
of thermal-fluids applications~\cite{nek5000}. 
%Nek5000 has scaled to millions of MPI ranks using 
%the Nek-based gsLib communication library to handle all near-neighbor and other 
%stencil-type communications~\cite{fischer15}.
%On CPUs, tensor contractions constitute the principal computational kernel 
%(typically $>$90\% of the flops). These can be cast as small, dense matrix-matrix 
%products resulting in high performance with a minor amount of tuning.
%
It employs high-order spectral elements \cite{pat84} in which the solution, 
data, and test functions  are represented as {\it locally structured} $N$th-order 
tensor product polynomials on a set of $E$ {\it globally
unstructured} curvilinear hexahedral brick elements. In the following sections
we provide some details from the point of view of RANS and hybrid RANS-LES
implementation. For more details the reader is referred to
appendix~\ref{sec:methodology_appn} and other published works in literature \cite{fischer15,dfm02}.    

%The approach yields two
%principal benefits.  First, for smooth functions such as solutions to the
%incompressible Navier--Stokes equations, high-order polynomial expansions
%yield exponential convergence with approximation order, implying a significant
%reduction in the number of unknowns ($n \approx EN^3$) required to reach
%engineering tolerances.  Second, the locally structured forms permit local
%lexicographical ordering with minimal indirect addressing and, crucially, the
%use of tensor-product sum factorization to yield low $O(n)$ storage costs and
%$O(nN)$ work complexities~\cite{sao80}.  As we demonstrate, the leading-order
%$O(nN)$ work terms can be cast as small, dense matrix-matrix products (tensor
%contractions) with favorable $O(N)$ work-to-storage ratios (computational
%intensity)~\citep{dfm02}.

%---------------------------------------------------------------------%
\subsection{RANS and Hybrid RANS-LES formulations}
The Reynolds-averaged Navier--Stokes equations are
\begin{equation}
\label{RANS}
\begin{aligned}
 &\frac{\partial \bar{u_i}}{\partial x_i}=0,\\
 &\frac{\partial \bar{u}_i}{\partial t}+\bar{u}_j\frac{\partial \bar{u}_i}{\partial x_j}
 =-\frac{1}{\rho}\frac{\partial \bar{p}}{\partial x_i}
 +\nu\frac{\partial^2 \bar{u}_i}{\partial {x_j}^2}
 %+2\nu\bar{S}_{ij}
 -\frac{\partial}{\partial x_j}\left(\overline{u'_i u'_j}\right),
 \end{aligned}
\end{equation}
where the velocity $u_i$ is decomposed into its mean $(\bar{u}_i)$ and
fluctuating component $(u'_i)$.
Here $u_i$ is the $i$th component of the velocity vector, $\rho$ is the
density, $p$ is the fluid pressure, and $\nu$ is the fluid kinematic viscosity.
% and $S_{ij}= (\sfrac{1}{2}) \left( \sfrac{\partial{u_i}}{\partial
% x_j}+\sfrac{\partial{u_j}}{\partial x_i}\right)$ is the velocity strain rate tensor.
RANS equations are similar to Navier--Stokes equations except for the last term 
$\overline{u'_iu'_j}$, the \textit{Reynolds stresses}, which have to be
modeled to close the above equations.

Most RANS turbulence models are based on the
\textit{Boussinesq} assumption in which momentum transfer due to turbulent motion
is modeled by using an extra viscosity, called eddy viscosity~\cite{lars}. 
%The idea is analogous to the principle that
%momentum transfer due to molecular motion of the fluid can be modeled by using
%the molecular viscosity of the fluid.
%The eddy viscosity can be obtained either by solving an algebraic equation, 
%for example in mixing length models or by solving some additional transport 
%equations as in scalar eddy-viscosity models.
%Scalar eddy viscosity turbulence models, particularly the $k-\epsilon$ and
%$k-\omega$ variants \citep{wilcoxbook}, are immensely popular among industrial
%applications.  
In \nek{}, the eddy viscosity is obtained by using two-equation models of the
k-$\omega$ class. 
The implementation and performance of the k-$\omega$ class of models
within \nek{} are described in detail in the work by 
\etal{Tomboulides}~\citep{Tomboulides2018}, whereas the implementation and
testing of the k-$\tau$ version of the k-$\omega$ model are described in 
\etal{Tomboulides}~\citep{Tomboulides2024}.

%----------------------------------------------------------%
\subsection{Standard $k-\tau$ model and its SST formulation}

A major drawback of the models based on $\omega$ is that its asymptotic value at
the walls is singular and a ``sufficiently” large value for $\omega$ needs to be
precribed as the boundary condition. This leads to the solution being sensitive
to near-wall grid spacing. While an acceptable solution for low-order methods,
the excessive near-wall gradients lead to persistent numerical stability issues
in high-order codes. The problem of near-wall treatment has been critically
assessed in \citep{Speziale1992, Lloyd2020}.
In this study we use the $k-\tau$ version of the SST model, based on
the recently published work of \etal{Tomboulides}\citep{Tomboulides2024}. In the
$k-\tau$ model, one solves for the variable $\tau=1/\omega$, which represents the
time scale of turbulence, which, unlike $\omega$, remains finite on solid
boundaries. This formulation builds on the work in
\citep{Speziale1992, Benton1996}, and \citep{ Medic2006}.

The model equations are as follows:
%
%----------------------------------------------------------%
\begin{eqnarray}
    \frac{\partial \left(\rho k \right)}{\partial t} + \nabla \cdot (\rho k \vect{u}) &=& \nabla \cdot \Gamma_k \nabla k + P_k - \rho \beta^* \frac{k}{\tau}, \label{eqn:ktau_k} \\
    \frac{\partial \left(\rho \tau \right)}{\partial t} + \nabla \cdot (\rho \tau
\vect{u}) &=& \nabla \cdot \Gamma_\omega \nabla \tau - \alpha \frac{\tau}{k}P_k
+ \rho \beta - 2 \frac{\Gamma \omega}{\tau} \left( \nabla \tau \cdot \nabla \tau \right). \label{eqn:ktau_des_tau}
    \label{eqn:ktau_sst}
\end{eqnarray}
The last term in the $\tau$ equation was implemented in the form proposed in \citep{Kok2000} as
\begin{eqnarray}
    S_\tau = 2 \Gamma \omega \left( \nabla \tau \cdot \nabla \tau \right)/\tau = 8 \Gamma \omega \left( \nabla \tau^{1/2} \cdot \nabla \tau^{1/2} \right). \label{eqn:ktau_stau} 
\end{eqnarray}
Again, a blending function is introduced in the $\tau-$equation to arrive at the SST variant,
\begin{eqnarray}
%\frac{\partial(\rho k)}{\partial t}+\nabla \cdot(\rho k \vect{u}) &=& \nabla \cdot \Gamma_k \nabla k+P_k-\rho \frac{\beta^* k}{\tau} \\
%
\frac{\partial(\rho \tau)}{\partial t}+\nabla \cdot(\rho \tau \vect{u}) &=& \nabla \cdot \Gamma_\omega \nabla \tau-\alpha \frac{\rho}{\mu_t} \tau^2 P_k+\rho \beta-2 \frac{\Gamma_\omega}{\tau} \rho \nabla \tau \cdot \nabla \tau,  \nonumber \\
 &+& 2\left(1-F_1\right)  \rho \sigma_{\omega2} \tau(\nabla k \cdot \nabla \tau), 
\end{eqnarray}
and we define the kinematic eddy viscosity, closure coefficents, and auxiliary relations as
\begin{eqnarray}
\mu_t &=& \frac{\rho a_1 k}{\max \left(a_1 \tau^{-1}, F_2 S\right)}, \\
F_1 &=& \tanh \left(\operatorname{arg}{ }_1^4\right), \\
\arg_1 &=& \min \left[\max \left(\frac{\tau \sqrt{k}}{\beta^* d}, \frac{500 \nu \tau}{d^2}\right), \frac{4 \sigma_{\omega2} k}{CD_{k \omega} d^2}\right], \\
CD_{k \omega} &=& \max \left(-2 \sigma_{\omega2} \frac{\nabla k \cdot \nabla \tau}{\tau}, 10^{-10}\right), \\
F_2 &=& \tanh \left(\arg _2^2\right), \\
\arg _2 &=& \max \left(\frac{2 \tau \sqrt{k}}{\beta^* d}, \frac{500 \nu \tau}{d^2}\right),
\end{eqnarray}
where $d$ is the distance to the nearest wall.  The following model constants are used: 
\begin{eqnarray}
  \alpha_1 =5/9, \beta_1 =0.075, \sigma_{k1} =0.85, \sigma_{\omega1} =0.5, \nonumber \\
  \alpha_2 = 0.44, \beta_2 = 0.0828, \sigma_{k2} = 1, \sigma_{\omega2} = 0.856. \label{eqn:komega_ddes_const1}
\end{eqnarray}

Model constants are computed by a blend from the corresponding constants of the $k-\epsilon$ and $k-\omega$ models via $\alpha=\alpha_1 \cdot F1 + \alpha_2 \cdot (1-F1)$,  and so on.
Similar to the $k-\omega$ models \citep{Menter1994}, the $k-\tau$ SST model can show undesirable far-field numerical artifacts. These are due to the fact that the $\tau$ field can continue to grow outside the boundary layer where k is negligible.  In order to circumvent this problem, a limiter on eddy viscosity, originally proposed by Benton et al.~\citep{Benton1996}, has been implemented. Accordingly, the production, the dissipation, and the transport terms in Eq.~(\ref{eqn:ktau_sst}) are multiplied by a factor $\mu_t/R$, where $R$ is the limiter defined as
\begin{eqnarray}
  R = \max \left(10\mu, \mu_t \right),
\end{eqnarray}
where $\mu_t$ is the turbulent eddy viscosity given by $\mu_t = \rho k \tau$.

%----------------------------------------------------------%
\subsection{$k-\tau$ SST DDES formulation}
%----------------------------------------------------------%

The governing equation for $k$ for the SST-DDES formulation~\citep{Menter2003} is defined by
\begin{eqnarray}
  \frac{\partial \left(\rho k \right)}{\partial t} + \nabla \cdot (\rho k \vect{u}) &=& \nabla \cdot \Gamma_k \nabla k + P_k - \rho k^{3/2}/l_{DDES}, \label{eqn:komega_des_k}
    %
%    \frac{\partial \left(\rho \omega \right)}{\partial t} + \nabla \cdot (\rho \omega \vect{u}) &=& - \nabla \cdot \left[ \left( \mu + \sigma_\omega \mu_t \right) \nabla \omega \right] + \alpha \frac{\rho}{\mu_t} P_k - \beta \rho \omega^2 \nonumber \\
%     &+& 2(1-F1)\rho \sigma_{\omega 2} \frac{\nabla k \cdot \nabla \omega}{\omega} \label{eqn:komega_des_omega} \\
    %
%    \mu_t &=& \rho \frac{a_1 \cdot k}{\max \left( a_1 \cdot \omega, F2 \cdot S\right)} 
    \label{eqn:komega_ddes}
\end{eqnarray}
%where $F_1$ and $F_2$ are the blending functions given by,
%
%\begin{eqnarray}
%  F_1   &=& \tanh \left(arg_1^4\right)\\
%  arg_1 &=& \min \left( \max \left( \frac{\sqrt{k}}{C_\mu d}, \frac{500\nu}{d^2 \omega}\right), \frac{4\rho\sigma_{\omega 2} k}{CD_{k\omega} d^2} \right) \\
%  CD_{k\omega} &=& \max \left( 2\rho \sigma_{\omega 2} \frac{\nabla k \cdot \nabla \omega}{\omega}, 10^{-10} \right) \\
%  F_2   &=& \tanh \left( arg_2^2 \right) \\
%  arg_2 &=& \max \left( \frac{2\sqrt{k}}{C_\mu \omega d}, \frac{500 \nu}{d^2 \omega} \right)
%\end{eqnarray}
%
%
where the DDES length scale $(l_{DDES})$ is given as

\begin{eqnarray}
  l_{DDES} &=& l_{RANS} - f_d \max \left(0, l_{RANS} - l_{LES} \right), \\
  l_{LES} &=& C_{DES} h_{max}, \\
  l_{RANS} &=& \frac{\sqrt{k}\tau}{C_\mu}, \\
  C_{DES} &=& C_{DES1} \cdot F_1 + C_{DES2} \cdot \left( 1-F_1 \right).
\end{eqnarray}
Here $h_{max}$ is the maximum edge length of a spectral element divided by the
polynomial order $N$, and $f_d$ is the empirical blending function that is computed, following \citep{Strelets2001}, as
\begin{eqnarray}
  f_d &=& 1-\tanh \left[ \left( C_{d1} r_d\right) ^ {C_{d2}}\right], \\
  r_d &=& \frac{\nu_t + \nu}{\kappa^2 d^2 \sqrt{0.5 \left( S^2 + \Omega^2 \right)}}  .
\end{eqnarray}
Here, $S$ is the magnitude of the strain-rate tensor, and $\Omega$ is the magnitude of the vorticity tensor. 
The model constants are as follows:
\begin{eqnarray}
  C_\mu = 0.09, \; \kappa = 0.41, \; a_1 = 0.31, \label{eqn:komega_ddes_const0}\\
  C_{DES1} = 0.78, \; C_{DES2} = 0.61, \; C_{d1} = 20, \; C_{d2} = 3. 
\end{eqnarray}

Time integration in Nek5000 is based on a semi-implicit splitting scheme
using $k$th-order backward differences (BDF$k$) to approximate the time
derivative coupled with implicit treatment of the viscous and pressure
terms and $k$th-order extrapolation (EXT$k$) for the remaining advection and forcing
terms. We have used $k=3$ here to achieve third-order accurate integration in
time, combined with characteristics-based time-step estimator that permits us to
use a CFL number of 3.0 \citep{Maday1990}. The pressure Poisson solve is treated with GMRES using
$p$-multigrid as a preconditioner.  Details of the formulation can be found in
\citep{fischer04,fischer17,malachi2022a}.

%\section{Theoretical Results}

\section{Simulation Results \label{sec:results}}
%-----------------------------------------------------------------%
\nek{} has been the subject of extensive validation for resolved turbulent calculations 
such as DNS \citep{Hosseini2016,Ducoin2019} and wall-resolved LES 
\citep{Obabko2015,Vinuesa2017b,Tanarro2018,Kumar2023a} and also for RANS 
\citep{Bhushan2014,Shaver2020,Shaver2021,Tomboulides2024}.  
In the upcoming sections we report on our effort to validate the implementation of hybrid 
RANS-LES methodologies in \nek{}.  First, we use the results of NASA's turbulence modeling 
resource \citep{nasaTMR0012} to verify the $k-\tau$ SST RANS implementation for \naca{0012}.  
Next, we validate the hybrid RANS-LES formulation by comparing our results for \naca{0012} 
at \re{2\times 10^6} and $AoA$ in the range of $[0^o, 90^o]$ with DDES formulation 
with those obtained by \etal{Bidadi}~\cite{Bidadi2023}, who used a low-order finite-volume-based code.  
Then, we conduct additional simulations and present an explanation for the apparent differences between simulations and wind tunnel 
results.

%-----------------------------------------------------------------%
\begin{figure*}[!htb]
  \centering
  \includegraphics[width=0.3\textwidth]{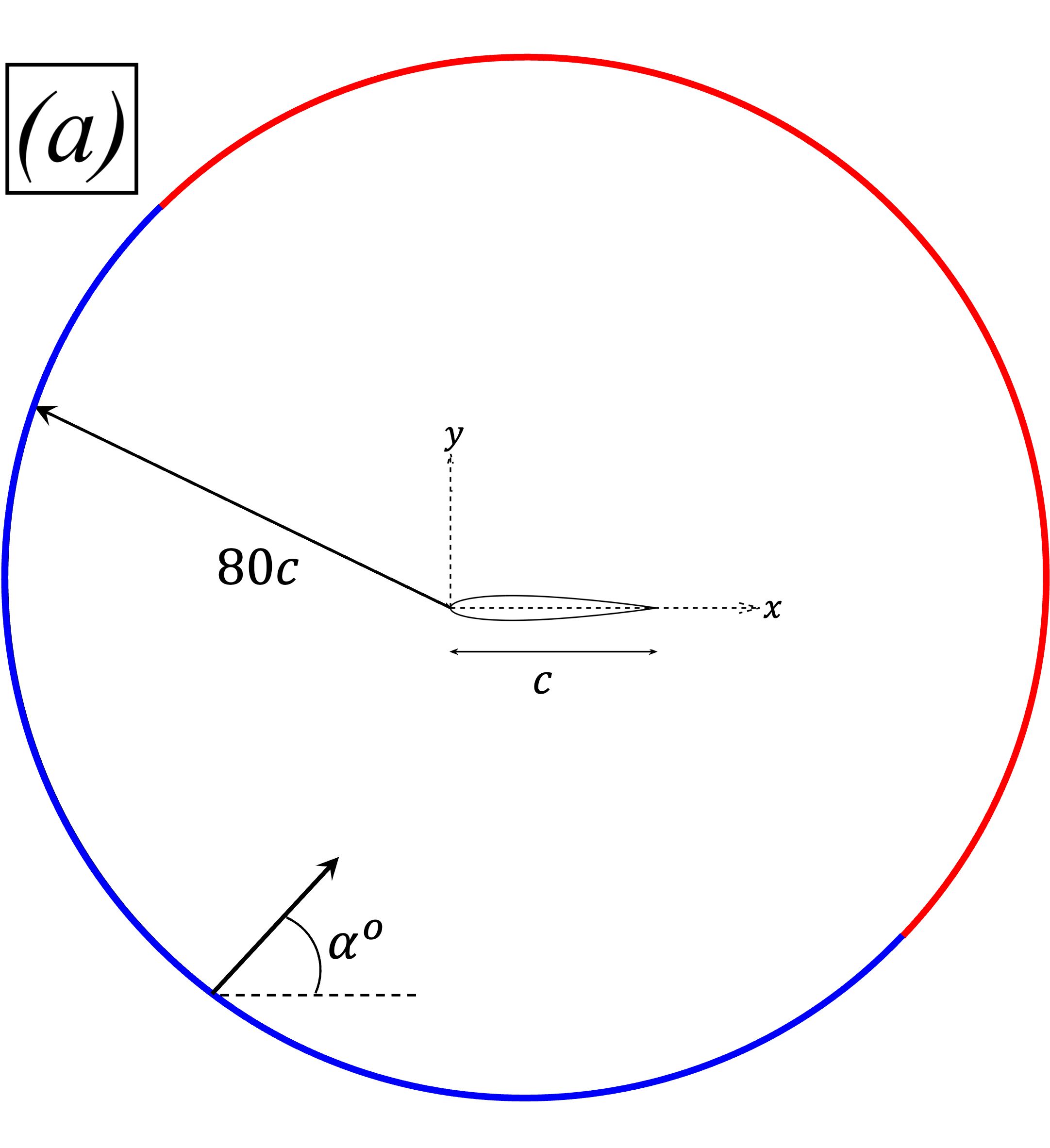}\quad%
  \includegraphics[width=0.3\textwidth]{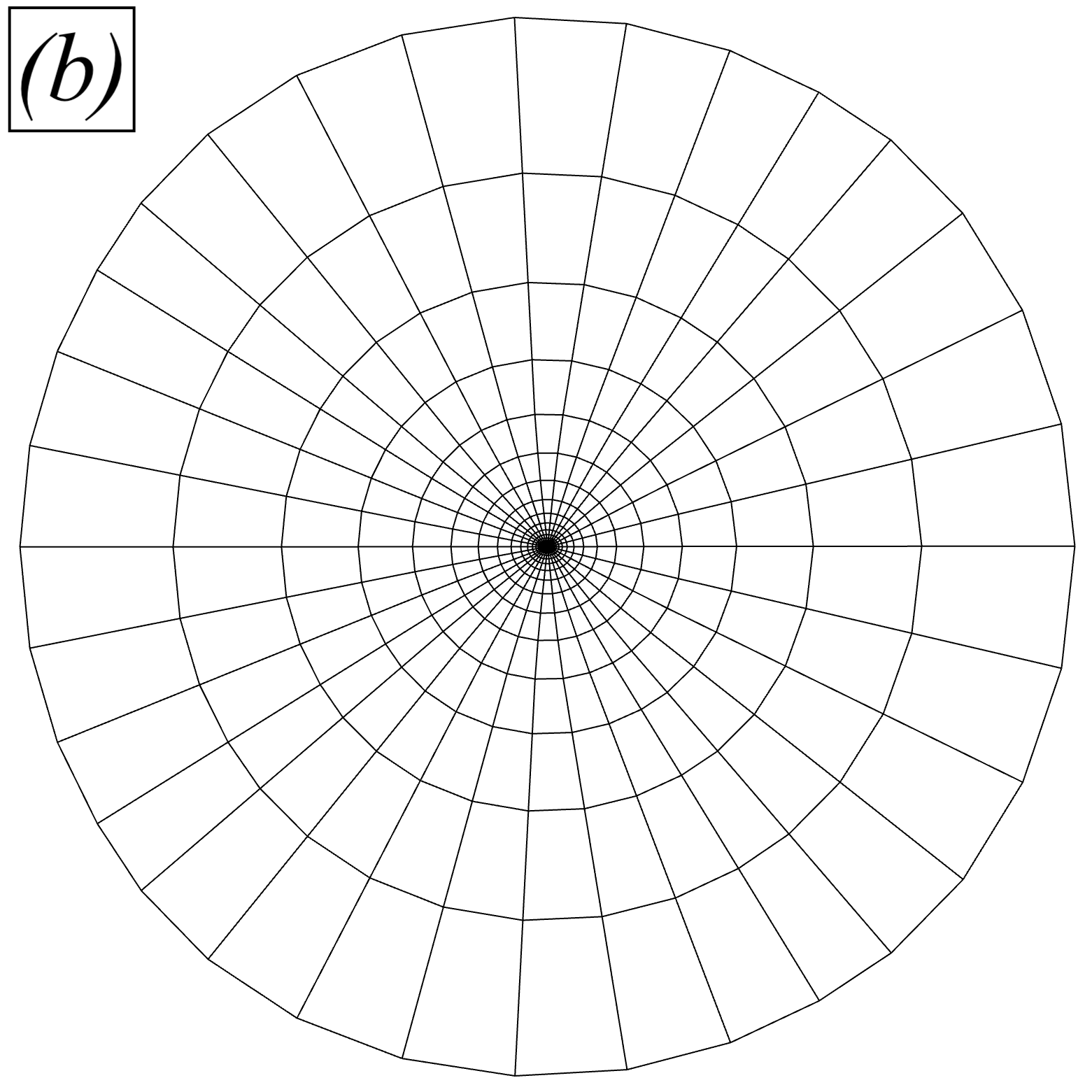}\quad% \\
  \includegraphics[width=0.3\textwidth]{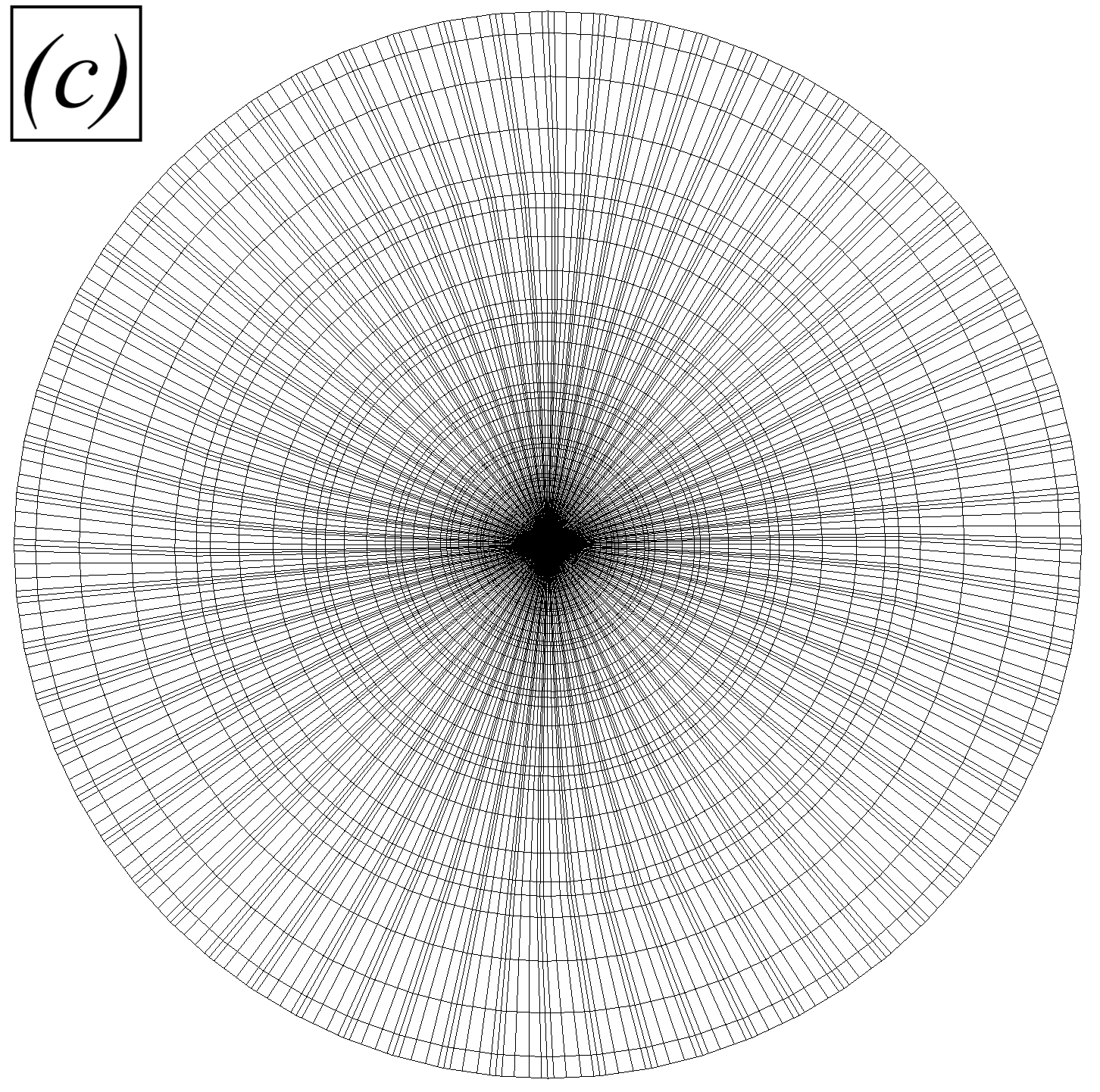}
  \includegraphics[width=0.3\textwidth]{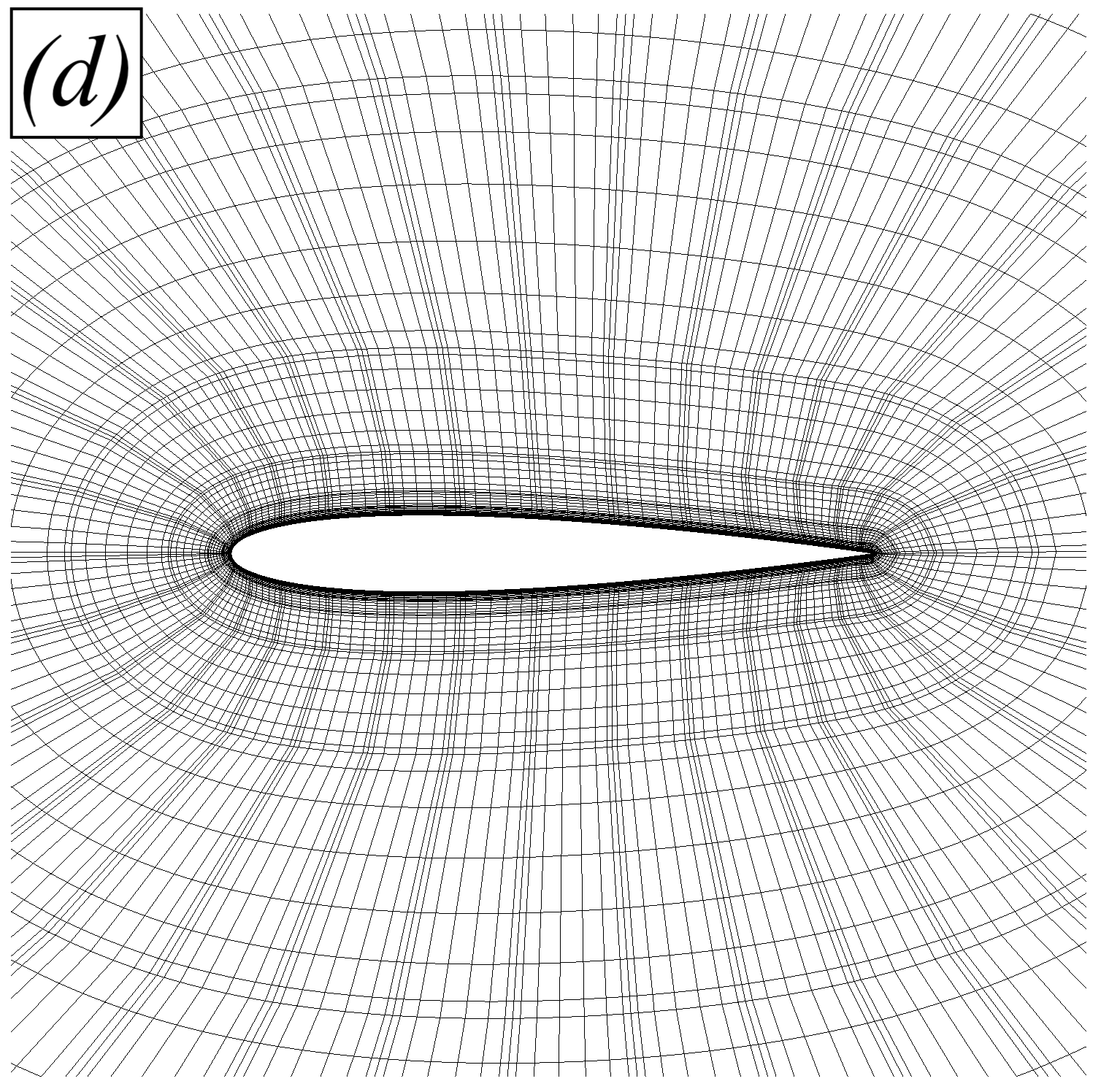}\quad%
  \includegraphics[width=0.3\textwidth]{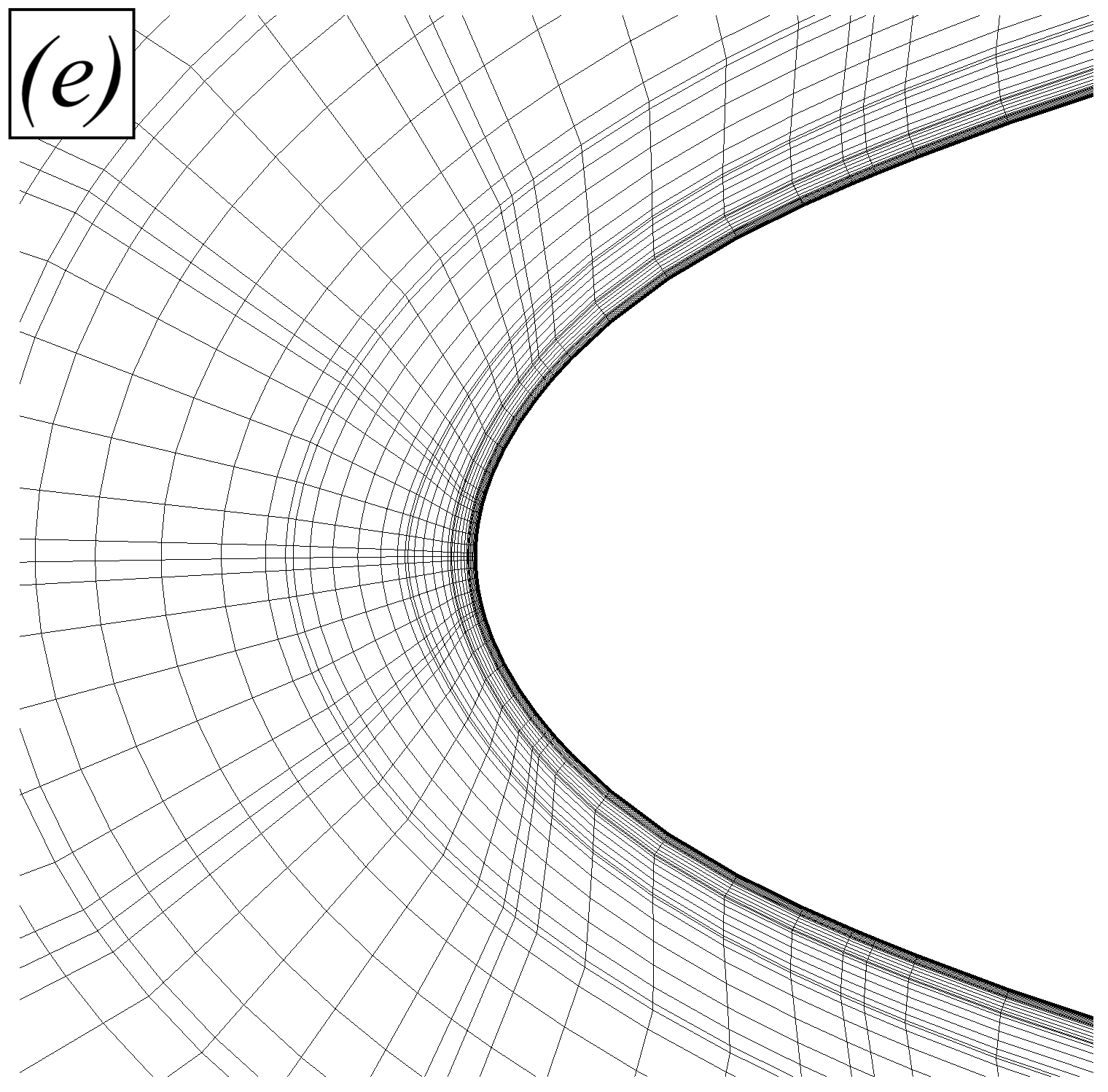}\quad%
  \includegraphics[width=0.3\textwidth]{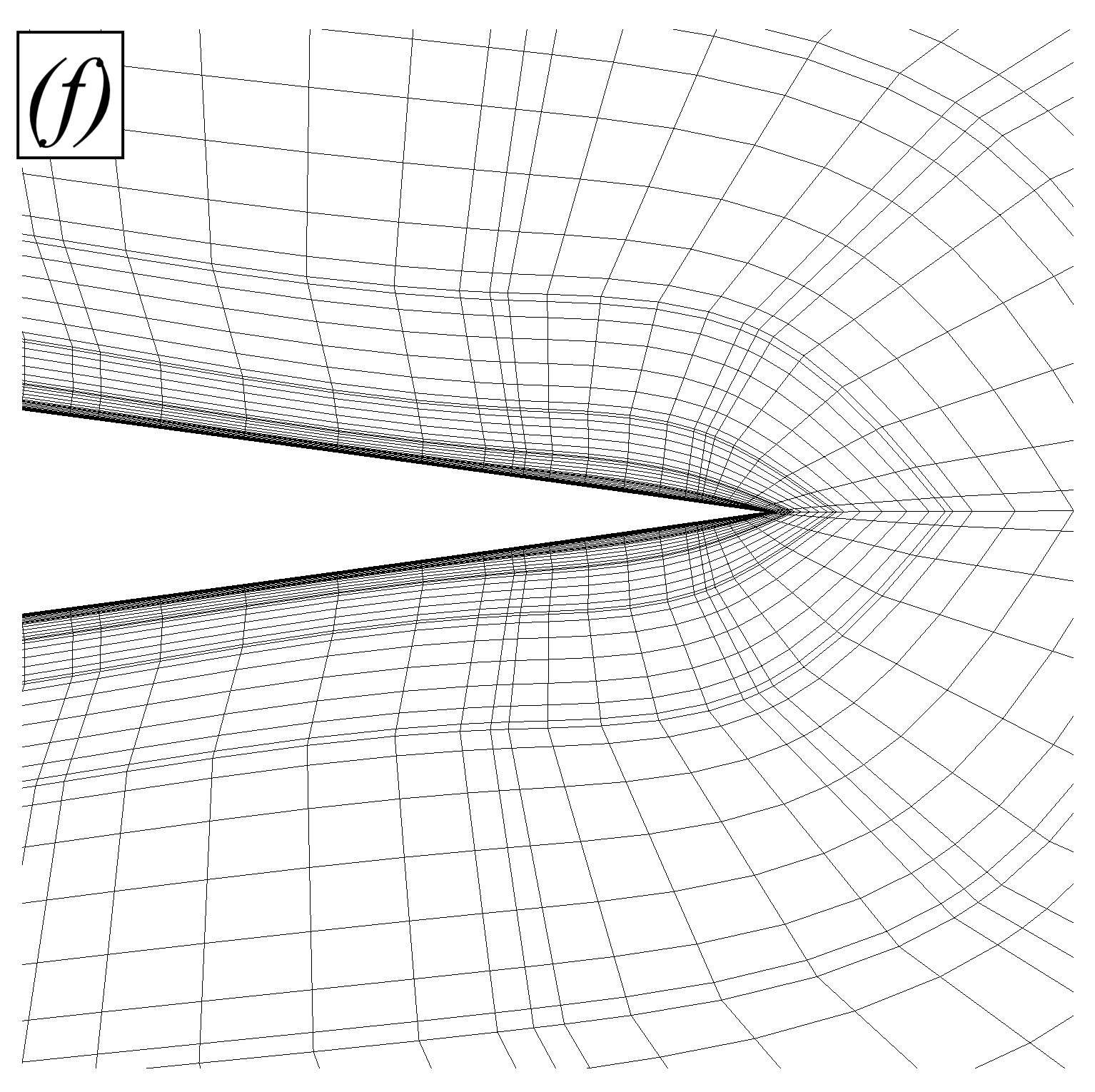}
  \caption{\label{fig:naca_o_grids} \textit{(a)} (not to scale) O-domain used for carrying out wing simulations; blue and red (based on incoming angle) represent inflow and outflow conditions, respectively; \textit{(b)} cross section of spectral element mesh used; \textit{(c)} sample grid with all GLL points; \textit{(d)} close-up of airfoil region; \textit{(e)} close-up of leading-edge region; and \textit{(f)} close-up of trailing-edge regions.}
\end{figure*}
%-----------------------------------------------------------------%

%-----------------------------------------------------------------%
\begin{figure*}[!htb]
  \centering
  \includegraphics[width=0.5\textwidth]{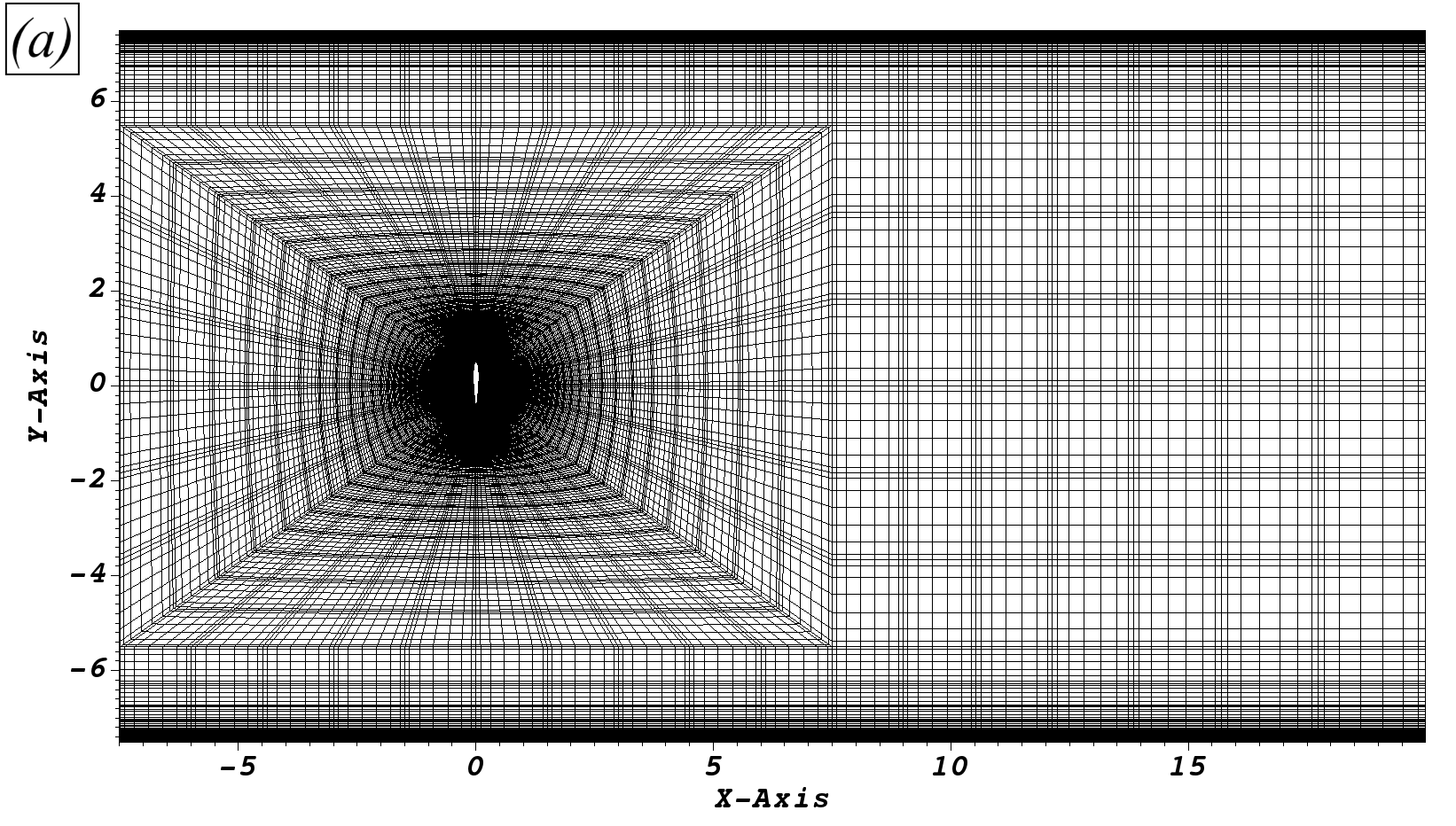}\quad%
  \includegraphics[width=0.3\textwidth]{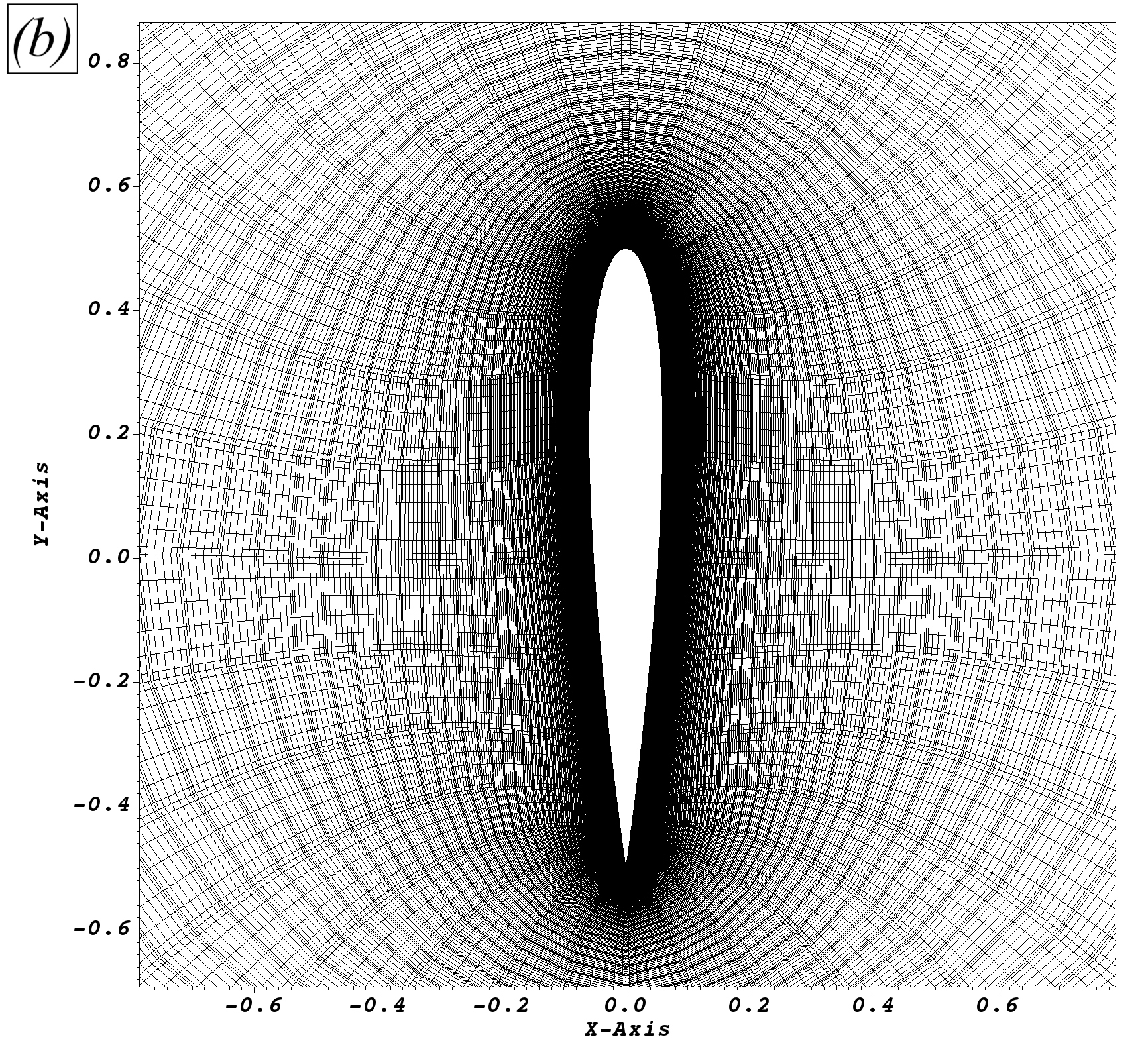}
  \caption{\label{fig:naca_t_grids} \textit{(a)} Cross-section view of mesh used for ``wind tunnel" wing simulations; \textit{(b)} close-up on airfoil. For resolution details refer to Table~\ref{table:ddes_re2m}.}
\label{fig:naca_tunnel}
\end{figure*}
%-----------------------------------------------------------------%

% -----------------------------------------------------------------%
\subsection{Grid generation and boundary conditions \label{sec:grid_gen}}
%-----------------------------------------------------------------%
%-----------------------------------------------------------------%
Unless otherwise mentioned, all the data reported for NACA simulations is
generated by using a O-type mesh, with a closed trailing-edge NACA profile.
Figure~\ref{fig:naca_o_grids} shows the domain and a sample grid used for the
present calculations.  A domain size of $80c$ in the $x-y$ plane is used
following other works in the literature.  For 3D calculations we have used a
spanwise domain size of $4$ times the chord length.  A no-slip boundary
condition is applied at the airfoil surface for $u_i$, $k$, and $\tau$. The
outer O-boundary is an inlet-outlet type, determined based on the angle that the
incoming flow makes with the outward-pointing surface normals. 

The generation of the O-mesh is a two-step process.  First, the 2D mesh
corresponding to the spectral elements was generated by using an open-source
grid generator, Construct2D \citep{construct2d}.  The elemental connectivity at
this stage is linear and results in an inaccurate approximation of the NACA
profile for low element count.  Mesh projecting and smoothing are required at
the final step to ensure an accurate representation of the wing geometry.  This
is done while importing the mesh in \nek{}.  We have used a third-order cubic
spline through the element vertices to interpolate the final
Gauss-Lobatto-Legendre (GLL) points used during the production runs.

In addition to unconfined simulations described above, we also carried out
confined simulations of certain NACA profiles in a ``numerical wind tunnel."
Figure~\ref{fig:naca_tunnel} shows the domain and a sample grid used for our confined 
simulation that we conducted at \aoa{90} configuration.  The dimensions of the tunnel 
are kept identical to the experiments of \cite{Critzos1955} 
($L_y \times L_z \equiv 15c \times 6c$). The streamwise dimension was extended
to apply an outlet-based boundary condition that accelerated the flow within the last
element on the right boundary. The flow acceleration helps avoid numerical
instabilities due to interaction of convecting voritces with the outlet
boundary condition. The boundary on the left assumes uniform inflow.  No-slip wall
conditions are applied both on the airfoil surface and on the side walls,
matching the setup employed in experimental wind tunnel studies.  
Grids for the confined case are designed using the meshing software 
Pointwise \cite{Pointwise}. Mesh projection and smoothing,
as discussed in the previous paragraph, is employed before the production runs. 
The $x-y$ resolutions are kept similar to that with the unconfined
simulation. The side wall boundary layers are not resolved since the intention
was to capture only the blocking effect.

%-----------------------------------------------------------------%
\begin{figure*}[!htb]
  \centering
  \includegraphics[width=0.8\textwidth]{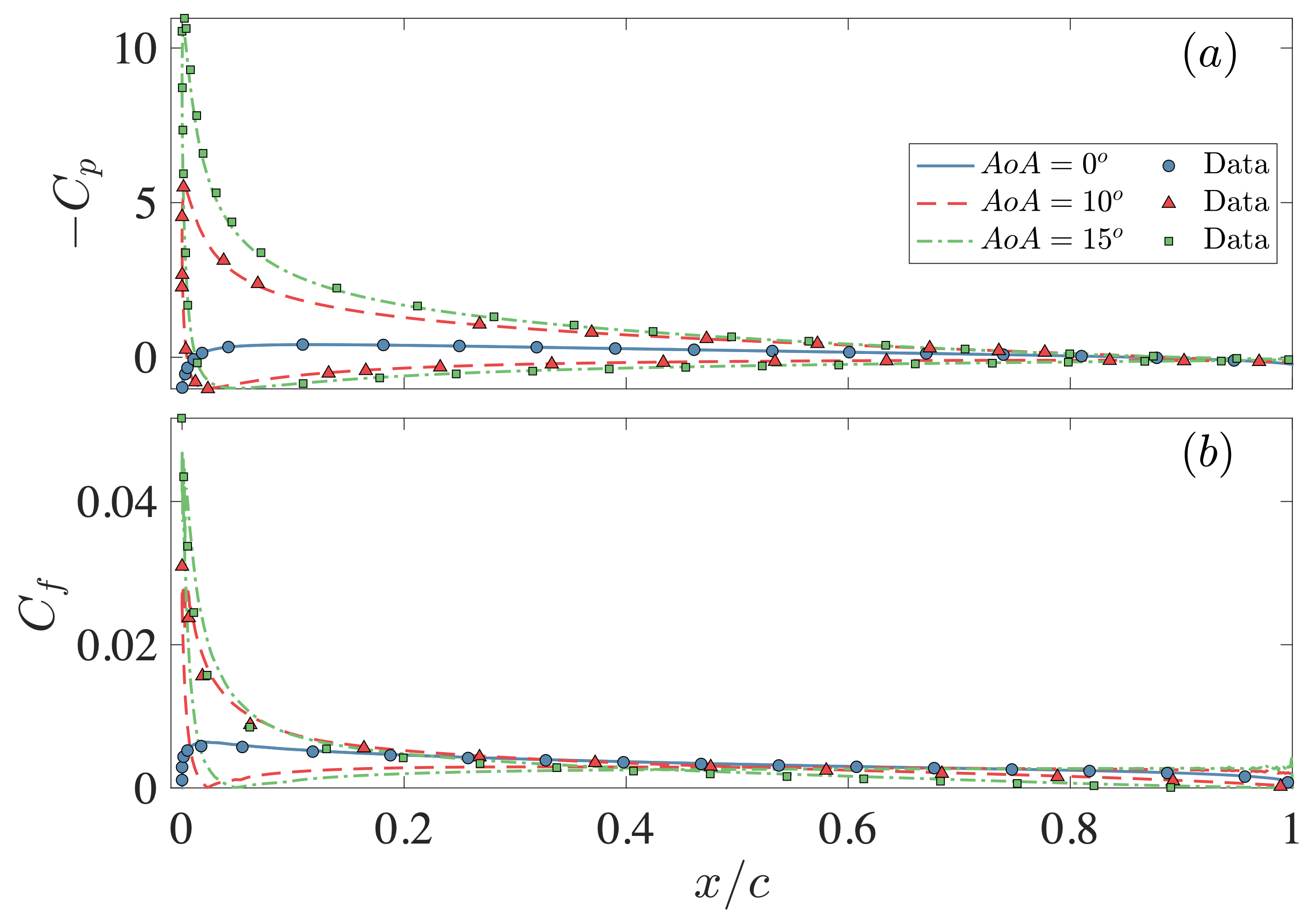}
  \caption{\textit{(a)} Pressure coefficient $(C_p)$ and \textit{(b)} friction
coefficient, $(C_f)$ for flow over \naca{0012} at \re{6\times 10^6} as predicted
by the $k-\tau$ SST model. ``Data" in the legend refers to data obtained with
NASA's CFL3D code. \label{fig:re6m_cpcf} }
\end{figure*}
%-----------------------------------------------------------------%

%-----------------------------------------------------------------%
% -----------------------------------------------------------------%
\subsection{RANS model validation \label{sec:rans_verify}}
%-----------------------------------------------------------------%
%-----------------------------------------------------------------%

We carried out a verification test of the RANS implementation in \nek{} using
the $k-\tau$ SST model for flow over 2D \naca{0012} profile at \re{6\times 10^6} and
\aoa{0,10,15}. This is a standard NASA benchmark test case for turbulence
modeling validation. Table~\ref{table:ktau_sst_6m} shows the parameters related
to the grid and the solver performance.  We have used a mesh resolution that is
on par with other such RANS studies. The performance of \nek{} solvers is also
good, with the highest angle taking around 10 GMRES iterations for pressure.
For the results reported, we use a characteristics-based CFL value of 3.0,
which results in a much higher time advancement than it would normally permit.  
Figure~\ref{fig:re6m_cpcf} compares the pressure ($C_p$) and friction ($C_f$)
coefficients for the three angles with the data of \cite{nasaTMR0012}.  They are
defined as follows:
\begin{eqnarray}
C_p = \frac{p-p_o}{0.5 \rho_o U_o^2}; \; C_f = \frac{\tau_w}{0.5 \rho_o U_o^2}.
\end{eqnarray}

We observe good agreement for $C_p, C_f$ between our case and the data from
NASA's CFL3D code \citep{nasaTMR0012}, which used Menter's SST model
\citep{Menter1994}.  We also compared the lift ($C_l$) and drag ($C_d$)
coefficients with the CFL3D simulations and with the experiments of
\cite{Ladson1985,Ladson1988}; they are tabulated in
Table~\ref{table:ktau_sst_6m_clcd}.  The lift and drag forces are defined as,
\begin{eqnarray}
C_l = \frac{2\mathcal{L}}{\rho_o U_o^2 c}; \; C_d = \frac{2\mathcal{D}}{\rho_o U_o^2 c}.
\end{eqnarray}
where $\mathcal{D}$ \& $\mathcal{L}$ are the forces, parallel and perpendicular
to the incoming flow respectively, integrated over the airfoil surface.
We observe an excellent match with the
results from CFL3D, with less than 1\% in lift and $<3\%$ difference in drag.  
The difference with the experiments, however, is larger, particularly for the drag
coefficient. This difference can be attributed to the limitations of the model
especially at higher values of AoA. 

%-----------------------------------------------------------------%
\begin{table}[!htb]
    \centering
    \begin{tabular}{ c c  c  c |  c  c}
        \toprule
        Case & $E$ & $N$ & $N_{dof} (M)$  & $\Delta t U_o/c$ & $n_{iter}$ \\ 
        \hline
        \aoa{0} & {\multirow{3}{*}{$64 \times 32$}}& {\multirow{3}{*}{$3$}} & {\multirow{3}{*}{$.55$}} & $5.2\times 10^{-4}$ & 3.9\\
        \aoa{10} &  &  &  & $2.5\times 10^{-4}$ & 11.6\\
        \aoa{15} &  &  &  & $1.4\times 10^{-4}$ & 10.8\\
        \bottomrule
    \end{tabular}
\caption{Parameters used for RANS with the $k-\tau$ SST model at $Re_c = 6\times
10^6$ and \aoa{0, 10, 15} for flow over \naca{0012}. $E$ refers to the number of
spectral elements used; $N$ refers to the polynomial order used inside the
spectral elements; $N_{dof}$ is the total number of degrees of freedom (in
million, $M$); $\Delta n^+$ denotes the wall-normal spacing in viscous unit;
$\Delta t$ is the average simulation time step at constant $CFL=3.0$; and
$n_{iter}$ is the average GMRES iterations for pressure solution. }
\label{table:ktau_sst_6m}
\end{table}

%-----------------------------------------------------------------%
\begin{table}[!htb]
    \centering
    \begin{tabular}{ c c c c | c c c}
       \toprule
               & \multicolumn{3}{c|}{$C_l$} & \multicolumn{3}{c}{$C_d$} \\
         \cline{2-7}
        Source  & \aoa{0} & $10^o$  & $15^o$  & \aoa{0} & $10^o$ & $15^o$ \\ 
       \midrule
        Present & $\approx 10^{-5}$ & 1.09  & 1.51  & 0.0083 & 0.013 & 0.023 \\ 
        Exp.  & -0.00724 & 1.05  & 1.49  & 0.008 & 0.012 & 0.018 \\
        CFL3D & 0 & 1.08  & 1.51  & 0.008 & 0.0125 & 0.0222 \\ 
       \bottomrule
    \end{tabular}
\caption{Lift and drag coefficients for RANS with the $k-\tau$ SST model at
$Re_c = 6\times 10^6$ and \aoa{0, 10, 15} for flow over \naca{0012}; Exp. refers
to the wind tunnel experiments of \cite{Ladson1985}.}
\label{table:ktau_sst_6m_clcd}
\end{table}

% -----------------------------------------------------------------%
\subsection{Assessment of DDES statistical sample requirements \label{sec:stats_conv}}
\begin{figure*}[!htb]
  \centering
  \includegraphics[width=0.7\textwidth]{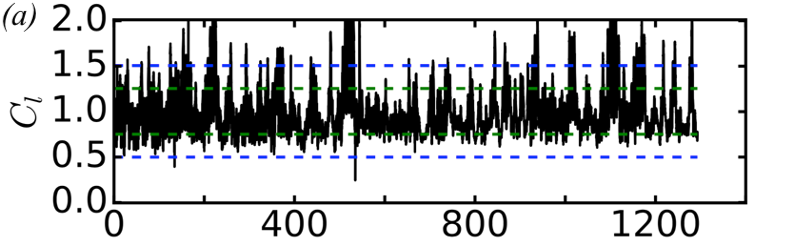}
  \includegraphics[width=0.7\textwidth]{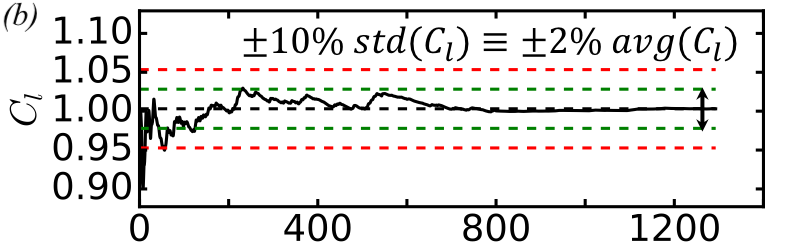}
  \includegraphics[width=0.6\textwidth]{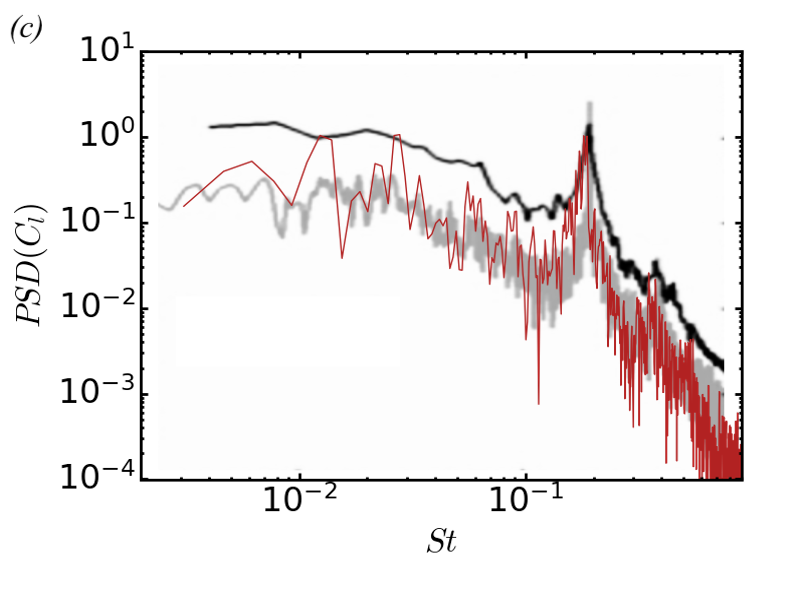}
  \includegraphics[width=1.0\textwidth]{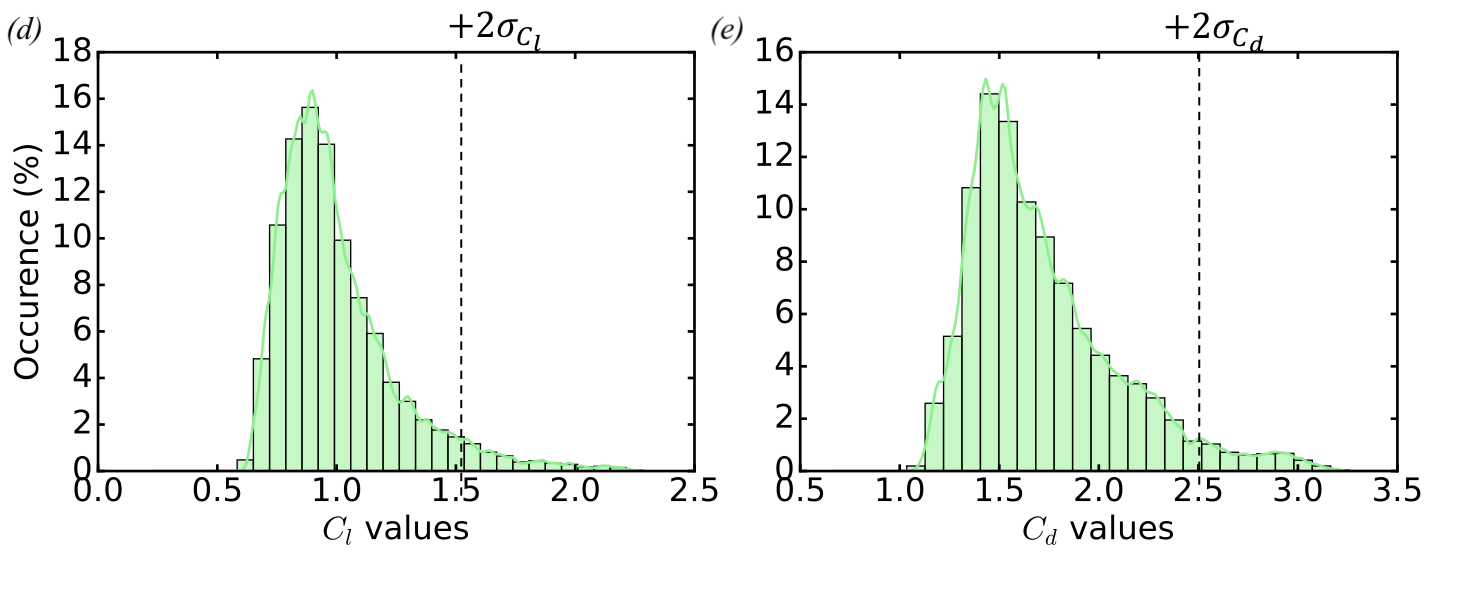}
  \caption{\textit{(a)} Time evolution of spanwise-averaged lift 
coefficient for \naca{0021} at \re{2.7\times10^5} and \aoa{60}; 
green and blue dashed lines depict $\pm \sigma(C_l)$ and $\pm 2 \sigma(C_l)$, 
respectively; 
\textit{(b)} running average of span-averaged $C_l$;
dashed-black line shows the sample mean; 
\textit{(c)} $C_l$ power spectral density; red line: \nek{}, 
black line: DES data of \cite{Garbaruk2009}; 
grey line: exp. data of \cite{Swalwell2003};
\textit{(d)} histogram (percent occurrence) of $C_l$; 
\textit{(e)} histogram (percent occurrence) of $C_d$.}
\label{fig:dt_dis1}
\end{figure*}

%-----------------------------------------------------------------%
Before assessing the accuracy of the DDES implementation in \nek{}, it is
instructive to quantify the statistical sample requirements vis-\`a-vis
recommendations for representative low-order codes mentioned by \etal{Garbaruk} 
\cite{Garbaruk2009} for flow over \naca{0021} at \re{2.7\times 10^5} and
\aoa{60}, simulated using a spanwise domain of $1c$.
Figure~\ref{fig:dt_dis1} (a,b) shows the span-averaged time history and the running
average of $C_l$ for \nek{} and compares it to \cite{Garbaruk2009}. 
We notice that it takes around 600 convective
time units for the running average (black dashed line) to converge within 0.2\% of the mean.  
This is about one-third of the time taken by fully implicit lower-order codes as
mentioned in \cite{Garbaruk2009}. In practice, such requirements for obtaining 
statistical convergence are seldom fulfilled by DES-based studies in literature. 
For example, in the review of \cite{Garbaruk2009} only one study satisfied this 
criterion. 
It was implicitly assumed that statistics collected over 500 convective times
was sufficient. Based on the data of \cite{Garbaruk2009}, this relaxed
criterion quantifies to an accuracy within $\pm 2\%$ of mean $C_l$. With
\nek{}, this is achieved in the first 100--150 convective time units (green dashed
lines in Fig.~\ref{fig:dt_dis1} (b)). 

Next, we analyze the effect of sample size on the accuracy of the
instantaneous results.
With an average $\Delta t \approx 7\times 10^{-5}$, we have used around 
 8.5M time samples. The fully implicit finite-volume data mentioned by 
\etal{Garbaruk}\cite{Garbaruk2009} used a time step size of 0.02, which 
corresponds to around $\approx 75,000$ samples.  
Figure~\ref{fig:dt_dis1} (c) shows the power spectral density of $C_l$
and compare it with the experimental data of \cite{Swalwell2003} as well. 
We have computed the spectrum using a fast Fourier transform of a uniformly
sampled signal, with a hanning window of size 2. 
We can observe that all three datasets in Fig.~\ref{fig:dt_dis1} (c)
yield the primary shedding frequency (at $St\approx 0.19$) and its  
harmonics in accordance with each other. Furthermore,  
our data is in good agreement with the experimental dataset across the entire
frequency range. The difference in power magnitude between our data and that of 
\etal{Garbaruk}\cite{Garbaruk2009} is particularly noticeable. They have 
 observed higher levels of fluctuations in $C_l$
(also noticeable clearly in Fig.~\ref{fig:dt_dis1} (b)), which we do not 
observe in \nek{}.  The source of these
fluctuations in the data of \cite{Garbaruk2009} is not clear at this point. We
remark, however, that they do not seem to stem from lower statistical
sampling rate.  We tested the \nek{} data with low sampling rates (results not
shown) but did not observe the same trend as that of \cite{Garbaruk2009}.  
The histograms of $C_l$ and $C_d$ are shown in
Fig.~\ref{fig:dt_dis1} (d,e). Our observed averaged $C_l$ and $C_d$ values are
1.0 and 1.72, respectively.  These are slightly higher than many studies mentioned
in \cite{Garbaruk2009}.  We notice in Fig.~\ref{fig:dt_dis1} (d,e) that the
distributions are skewed toward larger values.  It is likely that 
fully implicit methods, with much larger time steps, are prone to not capturing
the extreme events.    

%%-----------------------------------------------------------------%
%-----------------------------------------------------------------%
%-----------------------------------------------------------------%
%-----------------------------------------------------------------%

% -----------------------------------------------------------------%
\subsection{Assessment of DDES accuracy}
%-----------------------------------------------------------------%
In this section we comprehensively assess the accuracy of the DDES
formulation within \nek{}. For validation purposes we are following the work
of \cite{Bidadi2023} who conducted IDDES of \naca{0012} at \re{2\times 10^6}
and an angle of attack in the range of $0^\circ$--$90^\circ$. 
In addition, we also conducted several performance tests (not reported here) of
our DDES implementation in \nek{}. We observed excellent results in terms of
solver convergence and parallel performance, which ensures optimal operating
conditions for more complex studies to be undertaken in future. For more details
the reader is referred to our companion report \cite{Kumar2024}.

%% -----------------------------------------------------------------%
%\subsubsection{Grid refinement study \label{sec:mesh_reso}}
%%-----------------------------------------------------------------%

Table~\ref{table:ddes_re2m} shows the grid density used for our 3D simulations.
We conducted a grid refinement test using three different resolutions. 
The grid densities used for the grid-convergence tests are intentionally kept
nearly identical to those reported by \cite{Bidadi2023} for direct comparisons.  
The first GLL spacing for all the grids uses the same values of $\approx
1.2\times 10^{-5}$, matching that of \cite{Bidadi2023} as well.  This spacing
ensures $y^+\approx 1$ at $x/c=0.5$ for all angles considered in this study.

%-----------------------------------------------------------------%
\begin{table}[!htb]
    \centering
    \begin{tabular}{ c c  c  c  c c }
        \toprule
        Case & $N_{elm}$ & $N_{p}$ & $N_{dof} (M)$  & $\Delta n_{min}$ & $\Delta n^+$ \\ 
        \hline
%        2D        & $32 \times 14 $          & 7 & .29 & $1.2\times 10^{-5}$ & $<5$\\
%        \cline{2-5}
        Coarse & $32 \times 14 \times 5$  & 7 & 0.8 & $1.2\times 10^{-5}$ & $<5$\\
        Medium & $32 \times 20 \times 12$ & 7 & 2.6 & $1.2\times 10^{-5}$ & $<3$\\
        Fine   & $32 \times 35 \times 17$ & 7 & 6.5 & $1.2\times 10^{-5}$ & $<2$\\
        \cline{2-5}
        3D Tunnel & $3376 \times 9$ & 7 & 10.4 & $1.2\times 10^{-5}$ & $<5$
\footnotemark\\
        \bottomrule
    \end{tabular}
\caption{Simulation parameters used for DDES (with the  $k-\tau$ SST model) at
$Re_c = 2\times 10^6$ and \aoa{0--90} for flow over \naca{0012}. $N_{elm}$
refers to the number of spectral elements used; $N_p$ refers to the polynomial
order used inside the spectral elements; $N_{dof}$ is the total number of
degrees of freedom (in million, $M$); and $\Delta n^+$ denotes the wall-normal
spacing in viscous unit (for \aoa{30}).}
\label{table:ddes_re2m}
\end{table}
\footnotetext{resolution near airfoil surface only.}

%-----------------------------------------------------------------%
%%-----------------------------------------------------------------%
%\begin{table}[!htb]
%    \centering
%    \begin{tabular}{ c  c  c  c| c  c  c}
%        \toprule
%        $AoA$  
%	& $\Delta t \times 10^4$ & $n_{iter}$ & $t/t_{step} \times 10^3$ 
%	& $\Delta t \times 10^5$ & $n_{iter}$ & $t/t_{step} \times 10^1$\\ 
%        \hline
%            & \multicolumn{3}{c|}{2D Coarse} & \multicolumn{3}{c}{3D Coarse} \\
%            \cline{2-7}       
%            5  & 4.4  & 23 & 6.2 & 1.5 & 15 &  1.1 \\
%            17 & 0.75 & 31 & 7.3 & 2.1 & 66 &  1.8\\
%            30 & 0.31 & 32 & 7.2 & 5.4 & 88 &  2.1\\
%            45 & 0.34 & 35 & 7.5 & 2.7 & 58 &  1.8\\
%            60 & 0.26 & 34 & 7.2 & 1.9 & 53 &  1.7\\
%            90 & 0.21 & 30 & 7.0 & 1.5 & 40 &  1.5\\
%        \bottomrule
%    \end{tabular}
%\caption{Code performance for DDES with $k-\tau$ SST model at $Re_c = 6\times 10^6$ for flow over \naca{0012}. $\Delta t$ is the average simulation time step at constant $CFL=3.0$; $n_{iter}$ is the average GMRES iterations for pressure solution; and $t/t_{step}$ is the average CPU time per time step (in sec.) on 128 cores for 2D simulations and 256 cores for 3D simulations on LCRC's Improv cluster.}
%\label{table:ddes_3d_code_perform}
%\end{table}
%%-----------------------------------------------------------------%
%-----------------------------------------------------------------%
\begin{figure*}[!t]
  \centering
  \includegraphics[width=0.5\textwidth]{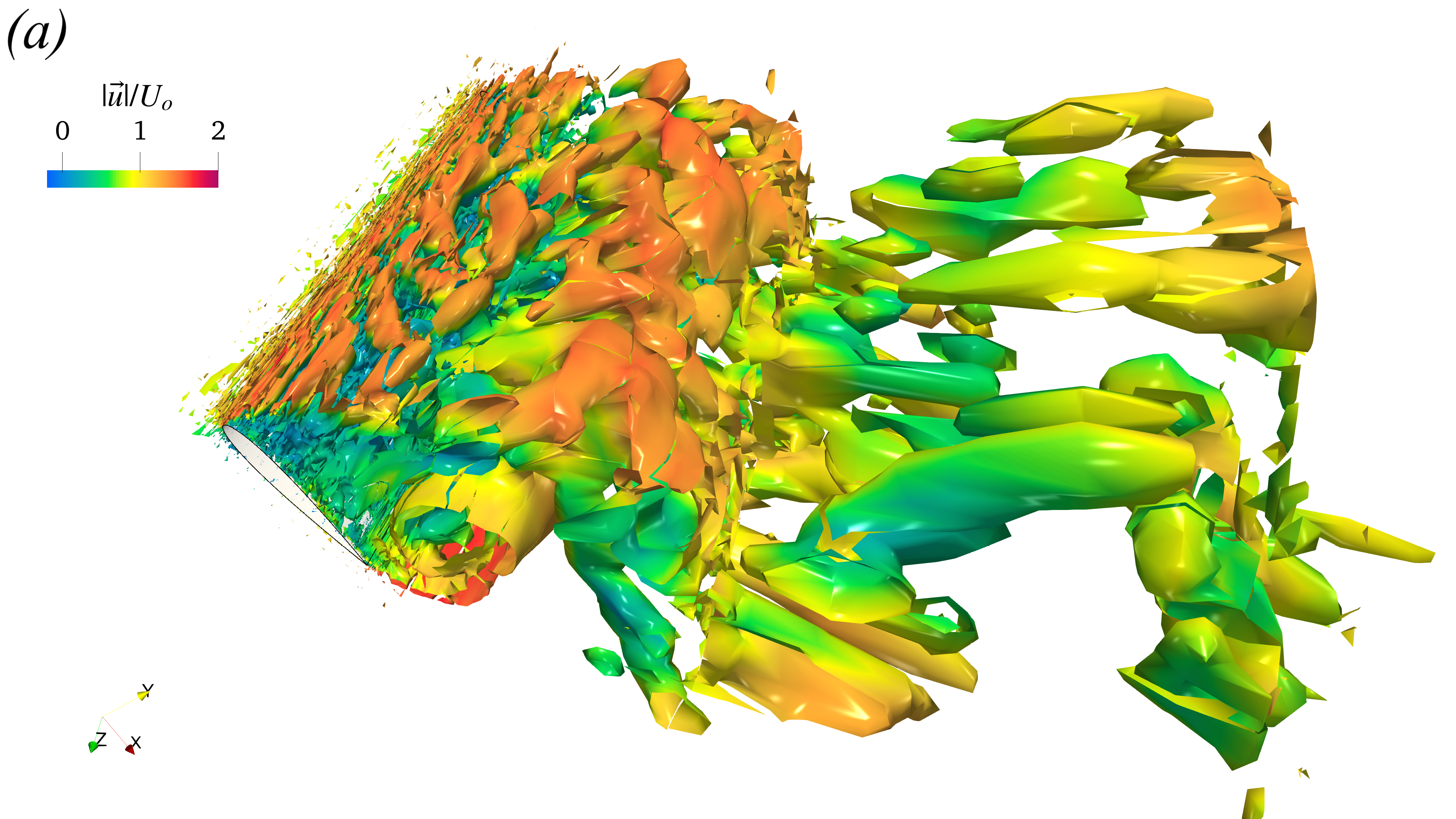}%
  \includegraphics[width=0.5\textwidth]{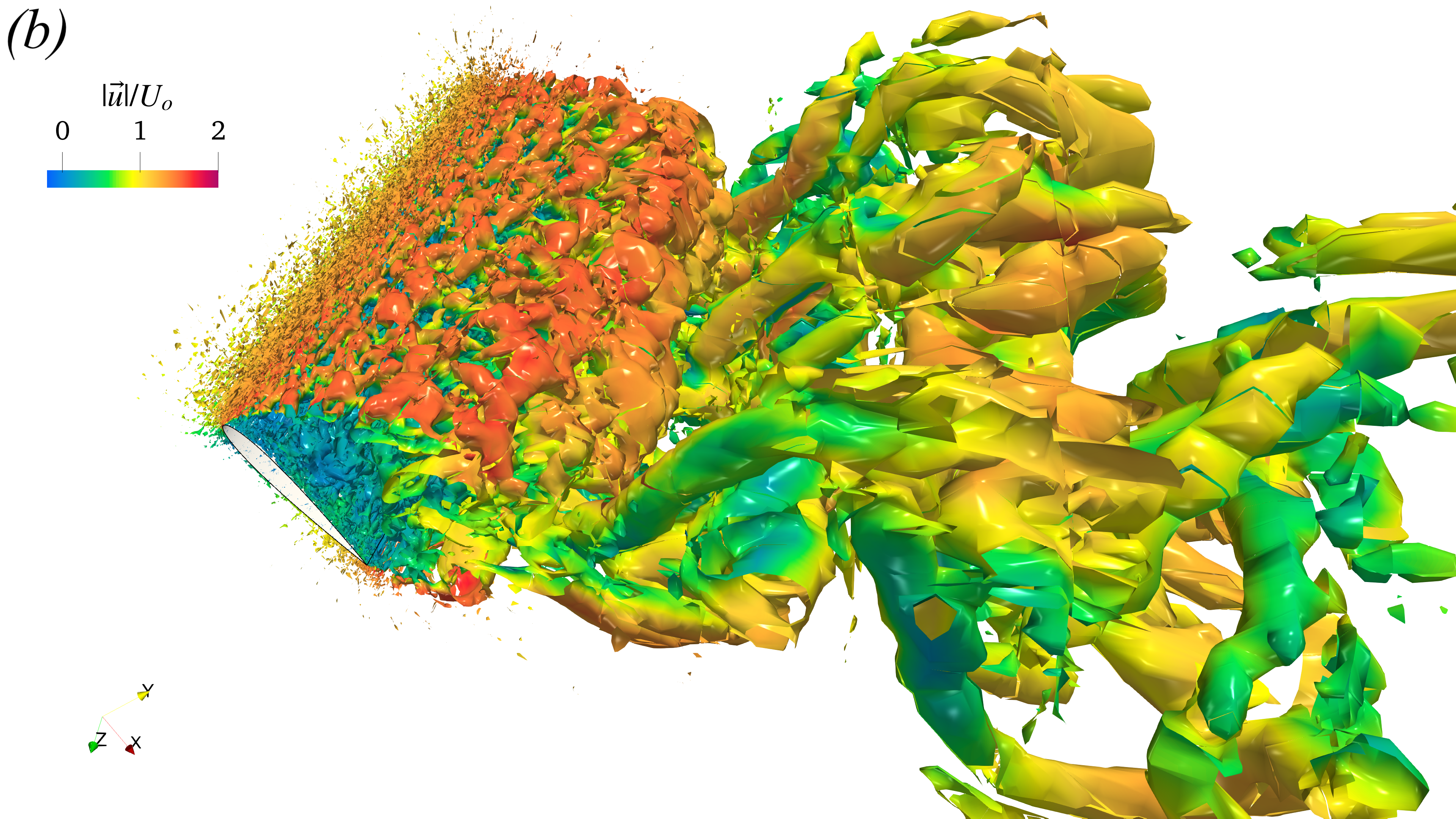}
  \caption{Visualization of vortical structures using contours of the Q-criterion
($Qc^2/U_o^2=1$) for the DDES cases at \aoa{45}; \textit{(a)} coarse grid;
\textit{(b)} Fine grid; contours are colored by velocity magnitude.}
  \label{fig:qcrit_ddes_3d}
\end{figure*}
%-----------------------------------------------------------------%
%-----------------------------------------------------------------%
\begin{table}[!htb]
    \centering
    \begin{tabular}{ c c }
               \large \smallskip $(\vec{u}\cdot \hat{\theta})/U_o$ & \large \smallskip $p/\rho U_o^2$ \\
       %--------%
       \raisebox{-.5\height}{{\large\textit{(a)} \smallskip}\includegraphics[width=0.25\textwidth]{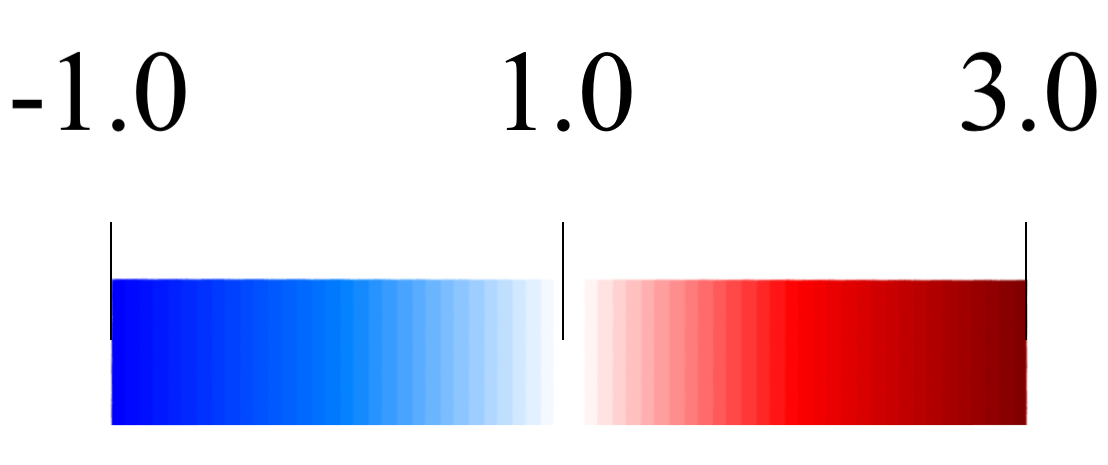}} & 
       \raisebox{-.5\height}{{\large\textit{(b)} \smallskip}\includegraphics[width=0.25\textwidth]{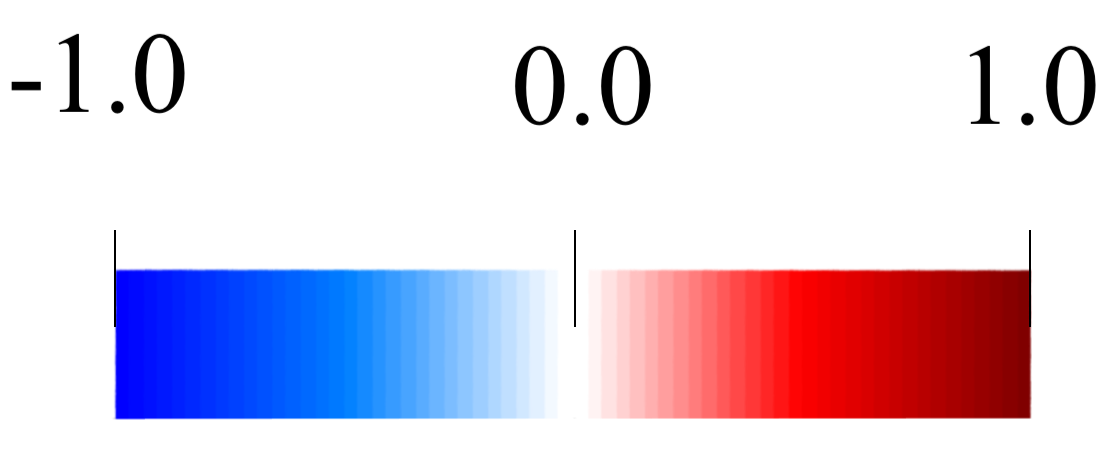}} \\ 

       %--------%
       \raisebox{-.5\height}{\includegraphics[width=0.4\textwidth]{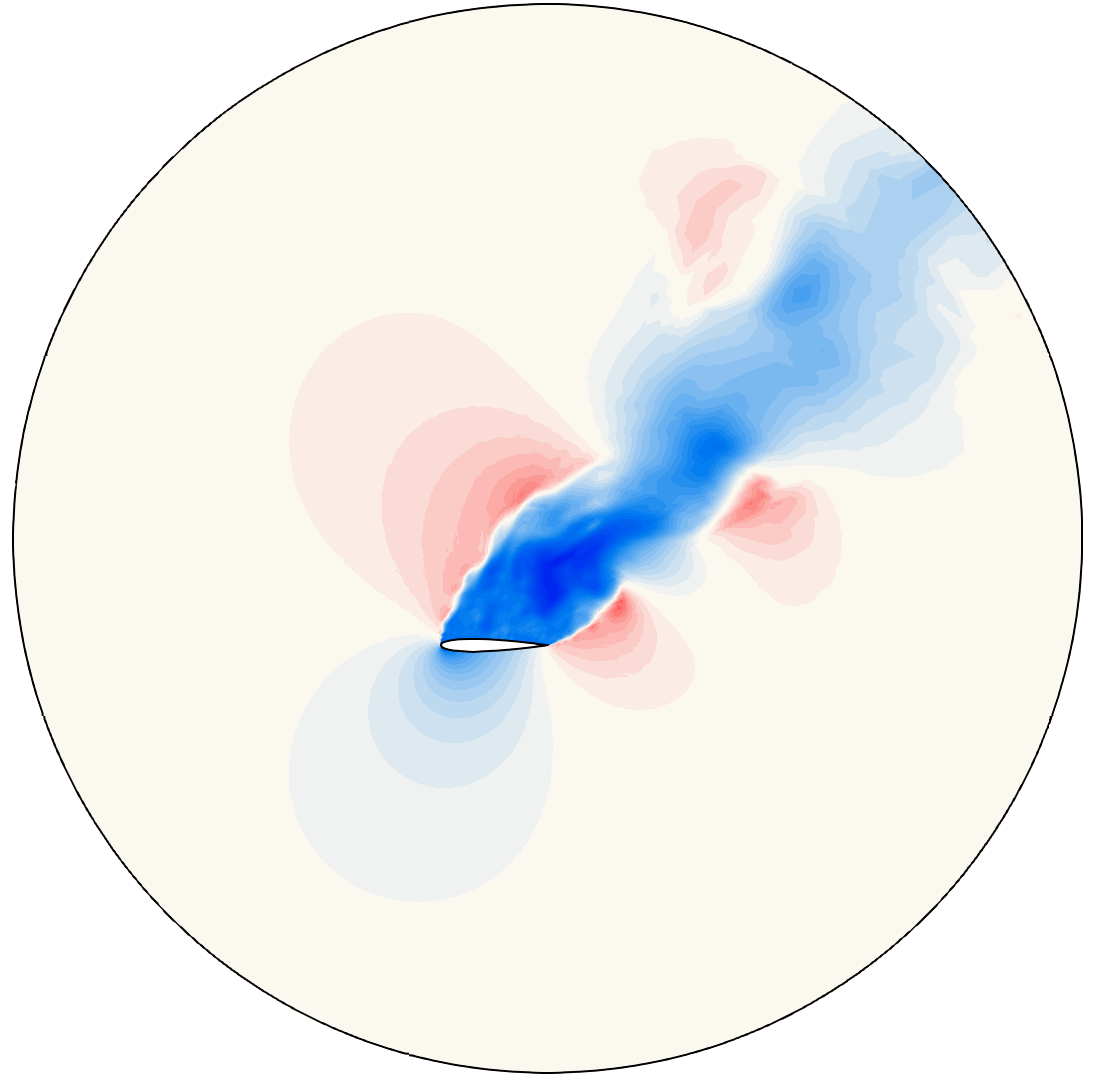}} & 
       \raisebox{-.5\height}{\includegraphics[width=0.4\textwidth]{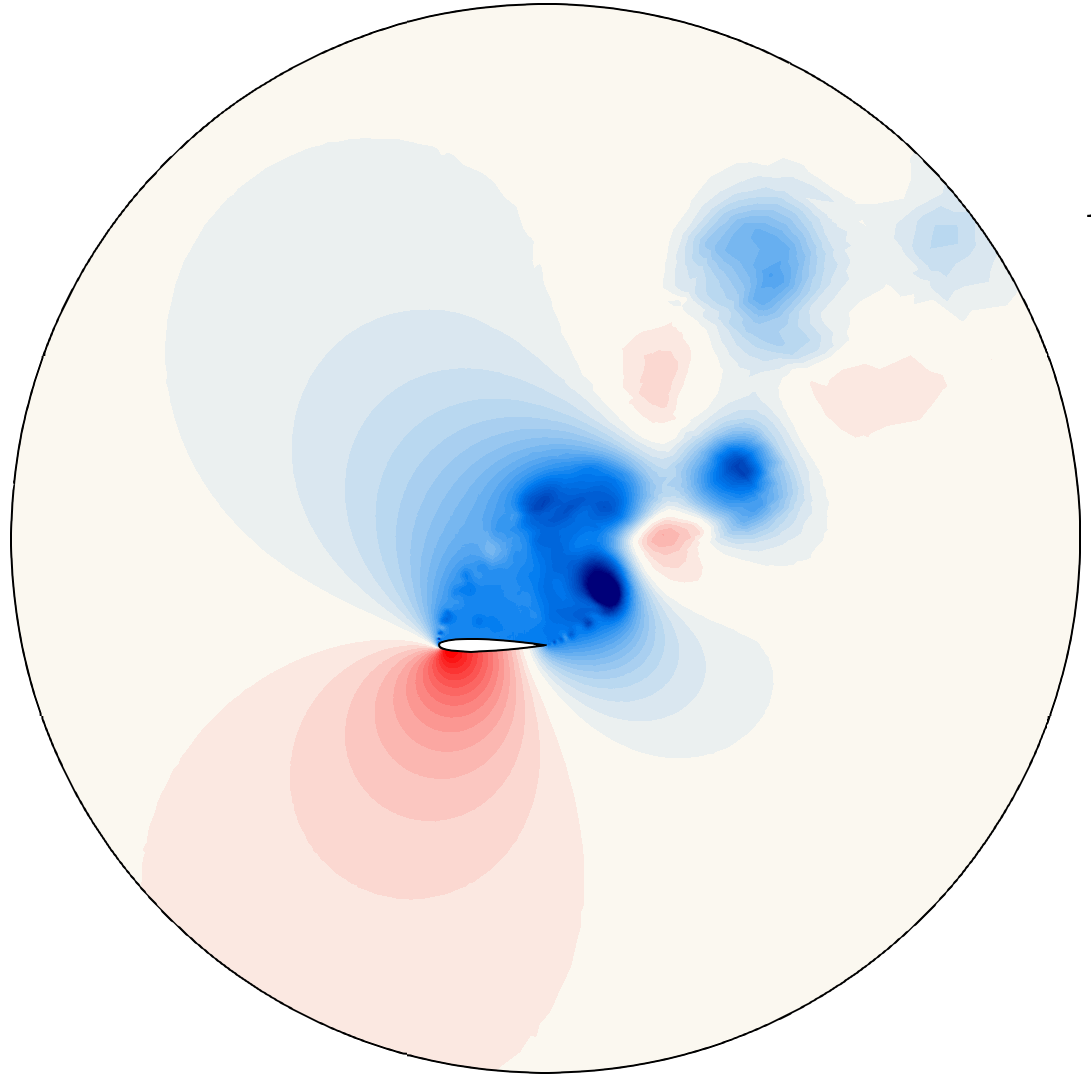}} \\

               \large \smallskip $\kappa/U_o^2$ & \large \smallskip $\tau U_o/c$ \\
       \raisebox{-.5\height}{{\large\textit{(c)} \smallskip}\includegraphics[width=0.25\textwidth]{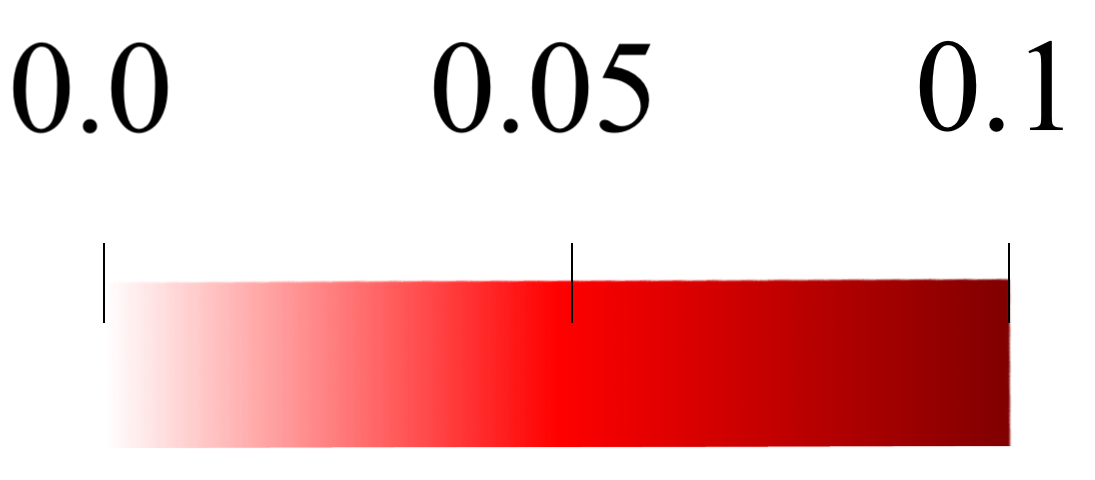}} &
       \raisebox{-.5\height}{{\large\textit{(d)} \smallskip}\includegraphics[width=0.25\textwidth]{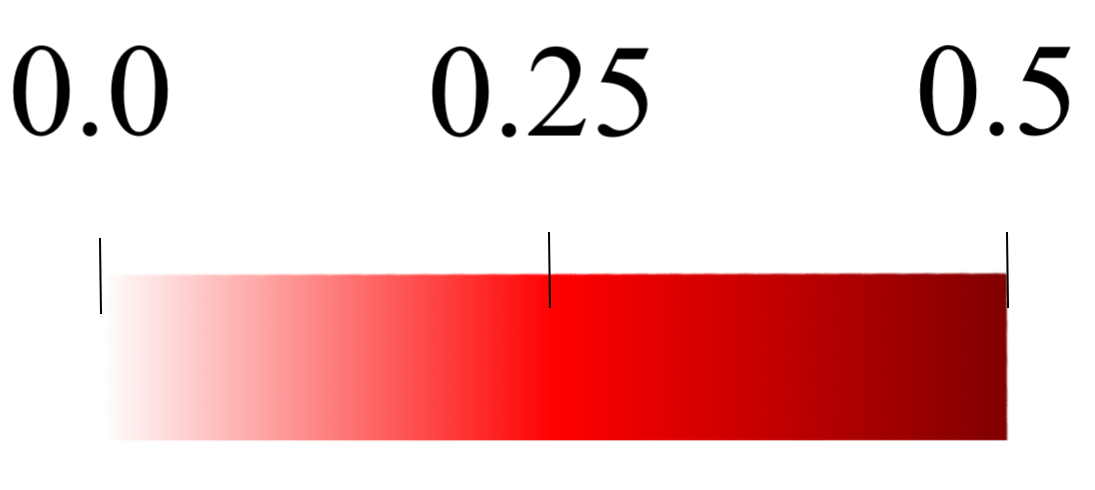}} \\
       %--------%
       \raisebox{-.5\height}{\includegraphics[width=0.4\textwidth]{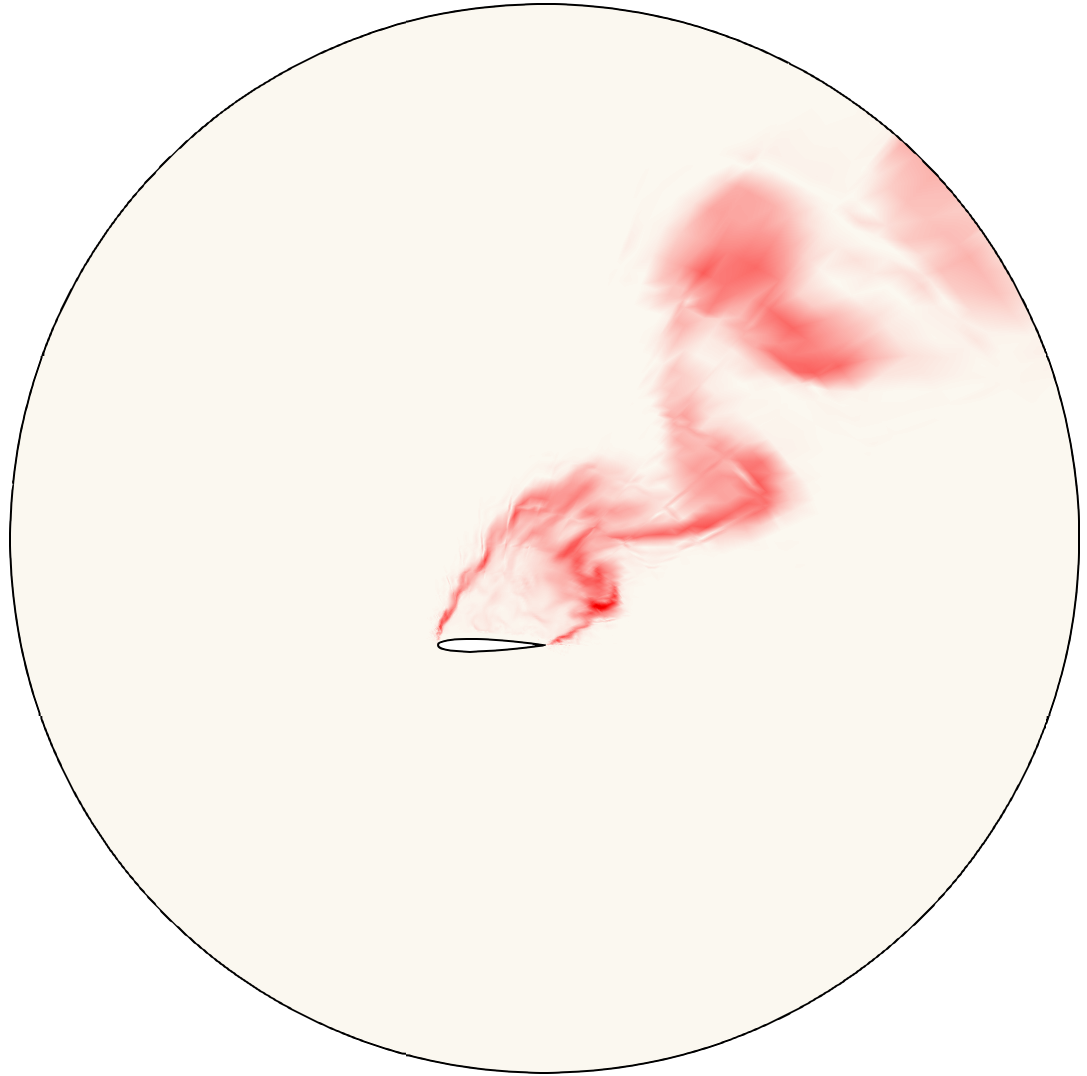}} &
       \raisebox{-.5\height}{\includegraphics[width=0.4\textwidth]{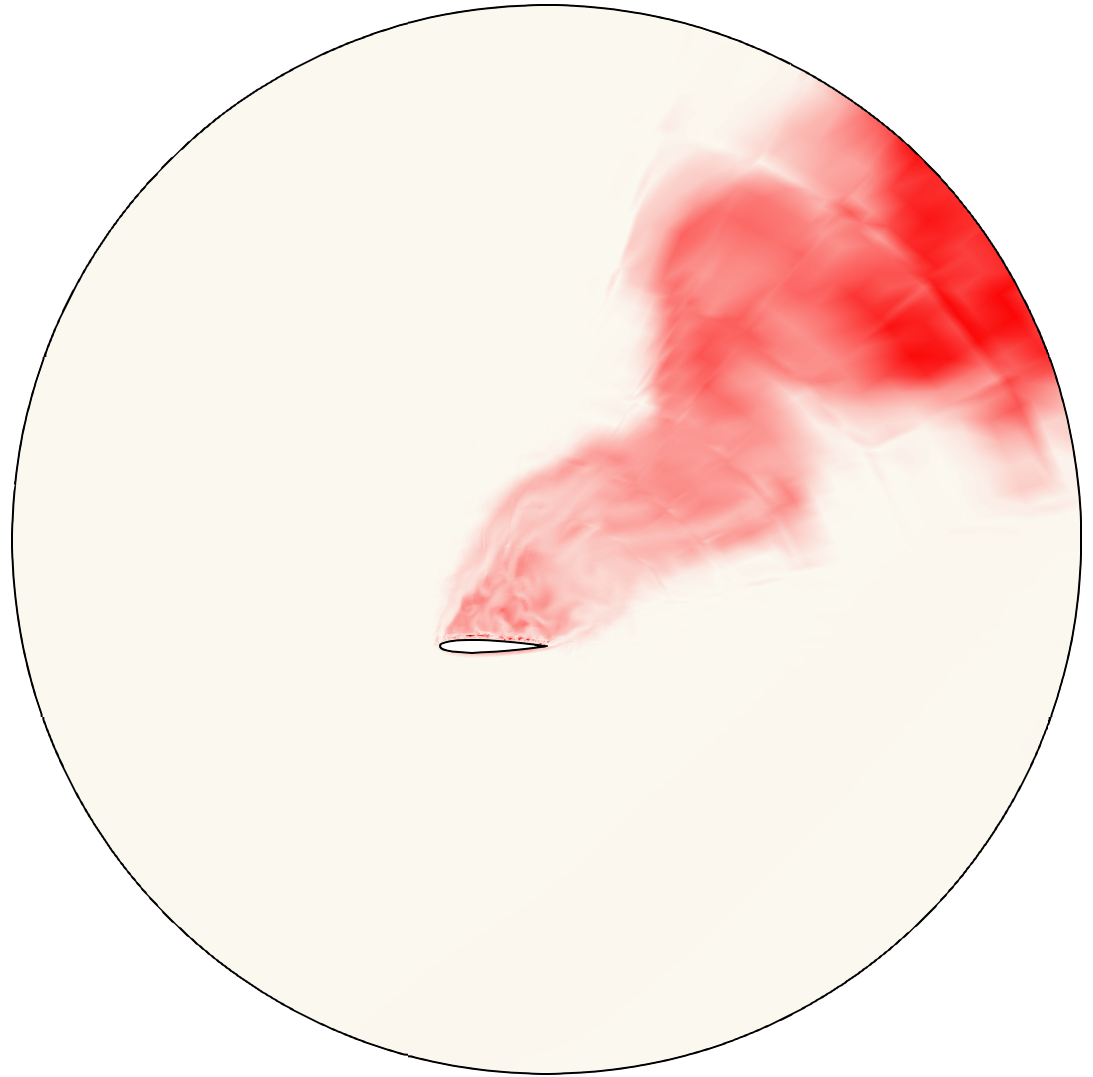}} \\
       %--------%
    \end{tabular}
\captionof{figure}{Visualization of instantaneous flow fields for unconfined flow over
\naca{0012} at $Re_c=2\times10^6$ and \aoa{45}; \textit{(a)} streamwise
velocity; \textit{(b)} pressure; \textit{(c)} model TKE; and \textit{(d)} model
time scale.}
\label{fig:naca_des_3d}
\end{table}
%-----------------------------------------------------------------%
%%-----------------------------------------------------------------%
\begin{table}[!htb]
    \centering
    \begin{tabular}{ c }
       %        \large $U/U_o$ & \large $\kappa/U_o^2$ & \large $\tau U_o/c$ \\
       %%--------%
       %\raisebox{-.5\height}{\includegraphics[width=0.2\textwidth]{./figs/TWing_03_01_00a}} & 
       %\raisebox{-.5\height}{\includegraphics[width=0.2\textwidth]{./figs/TWing_03_01_00c}} &
       %\raisebox{-.5\height}{\includegraphics[width=0.2\textwidth]{./figs/TWing_03_01_00d}} \\
       %%--------%
       %\raisebox{-.5\height}{\includegraphics[width=0.3\textwidth]{./figs/TWing_01_02a}} & 
       %\raisebox{-.5\height}{\includegraphics[width=0.3\textwidth]{./figs/TWing_01_02c}} &
       %\raisebox{-.5\height}{\includegraphics[width=0.3\textwidth]{./figs/TWing_01_02d}} \\
       %%--------%
       %\raisebox{-.5\height}{\includegraphics[width=0.3\textwidth]{./figs/TWing_01_03a}} & 
       %\raisebox{-.5\height}{\includegraphics[width=0.3\textwidth]{./figs/TWing_01_03c}} &
       %\raisebox{-.5\height}{\includegraphics[width=0.3\textwidth]{./figs/TWing_01_03d}} \\

       \large $(\vec{u}\cdot \hat{\theta})/U_o$ \hspace{2.5cm} \large $\nu_t/\nu \times 10^{-4}$ \\
       \includegraphics[width=0.7\textwidth]{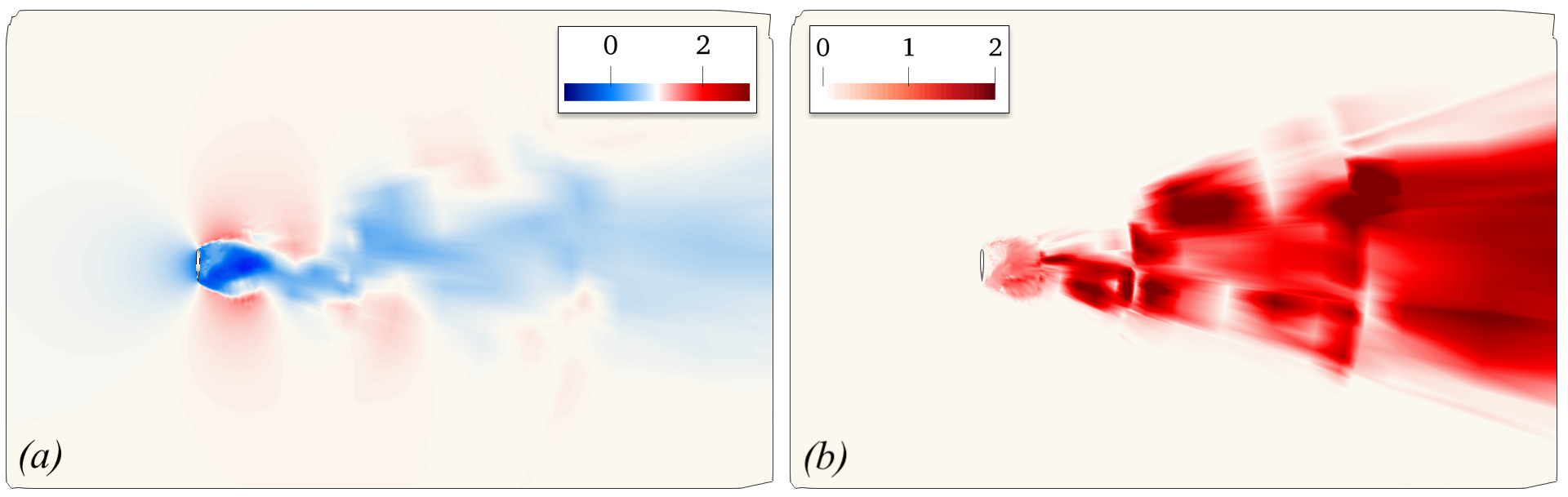} \\
       \includegraphics[width=0.7\textwidth]{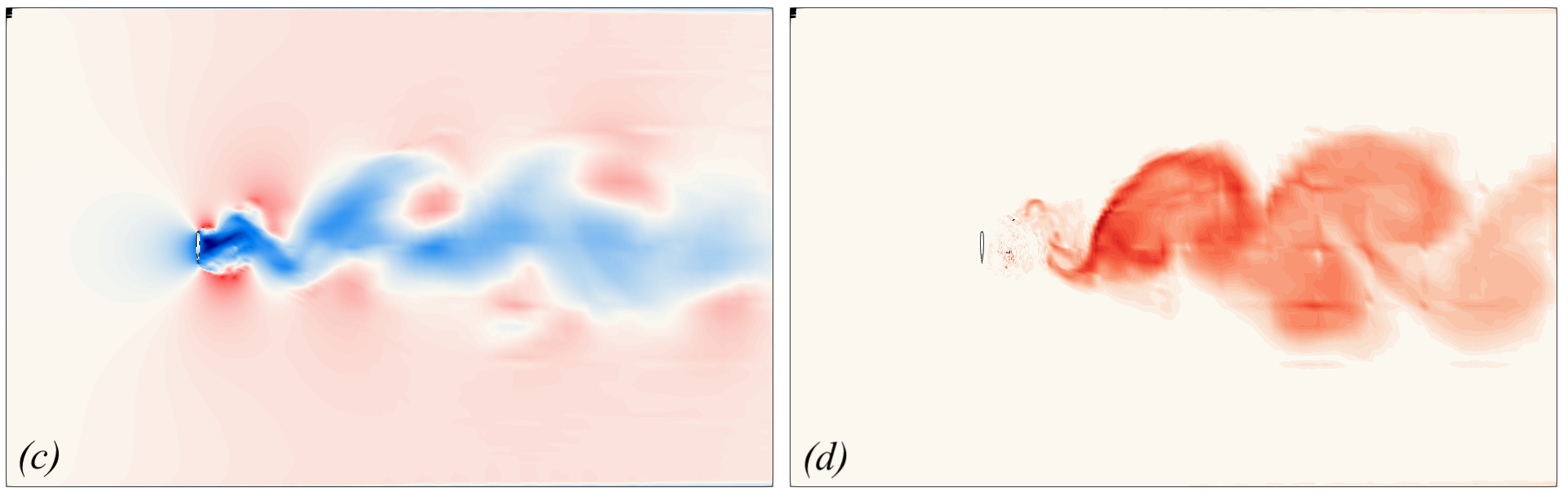}
    \end{tabular}
\captionof{figure}{Visualization of instantaneous flow fields for flow over
\naca{0012} at $Re_c=2\times10^6$ and \aoa{90}; \textit{(a,b)} unconfined
case; \textit{(c,d)} confined case; \textit{(first column)} streamwise velocity
and \textit{(second column)} turbulent eddy viscosity.}
\label{fig:naca_tunnel_views}
\end{table}

\begin{figure*}[!htb]
  \centering
  \includegraphics[width=0.3\textwidth]{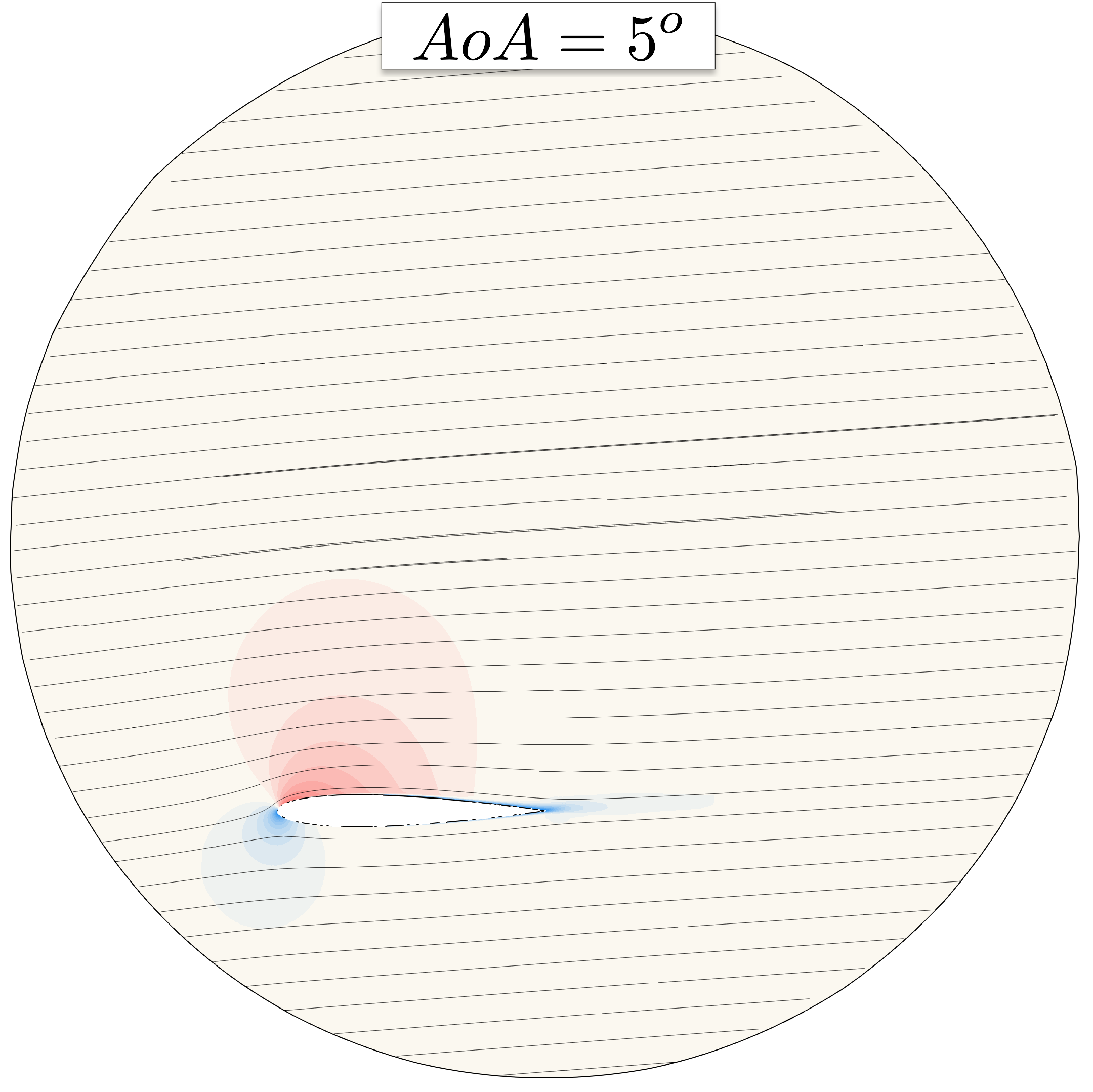}%
  \includegraphics[width=0.3\textwidth]{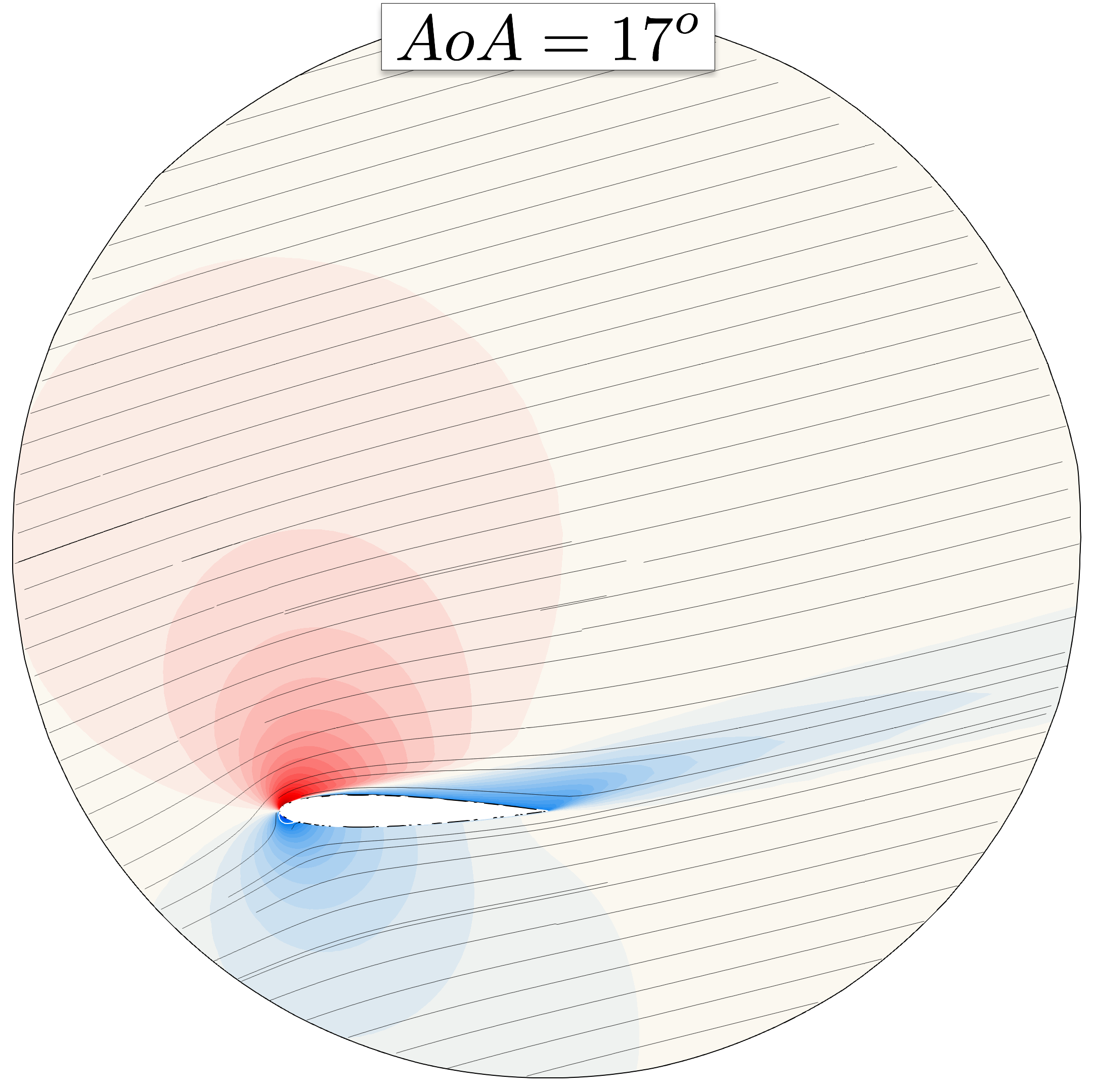}%
  \includegraphics[width=0.3\textwidth]{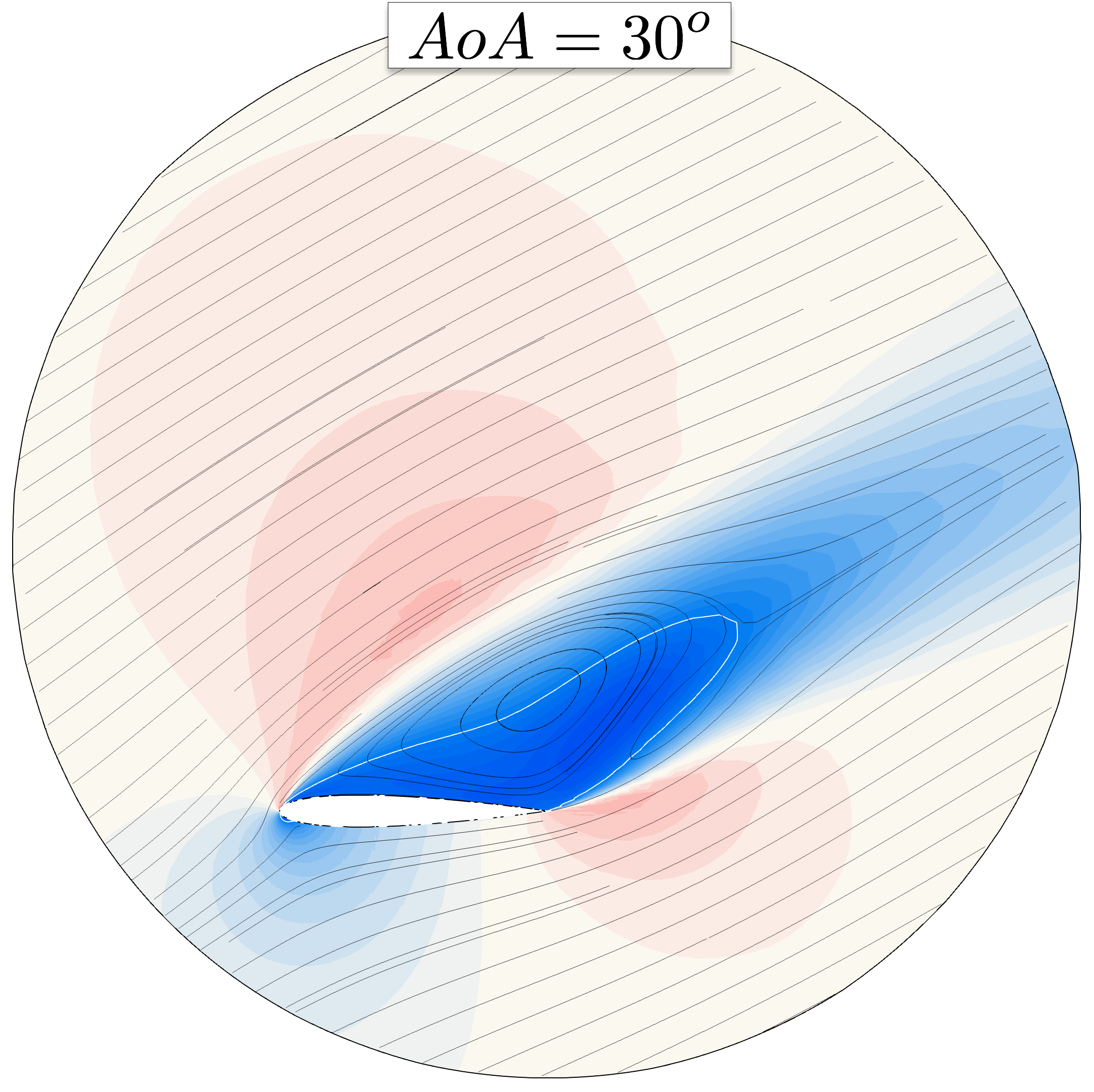}%
  \raisebox{.5\height}{\includegraphics[width=0.1\textwidth]{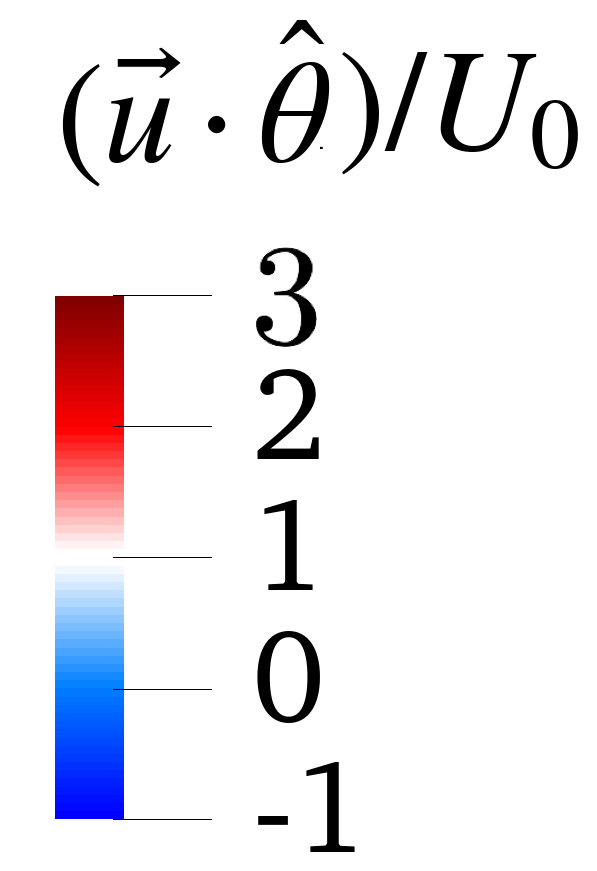}}
  \includegraphics[width=0.3\textwidth]{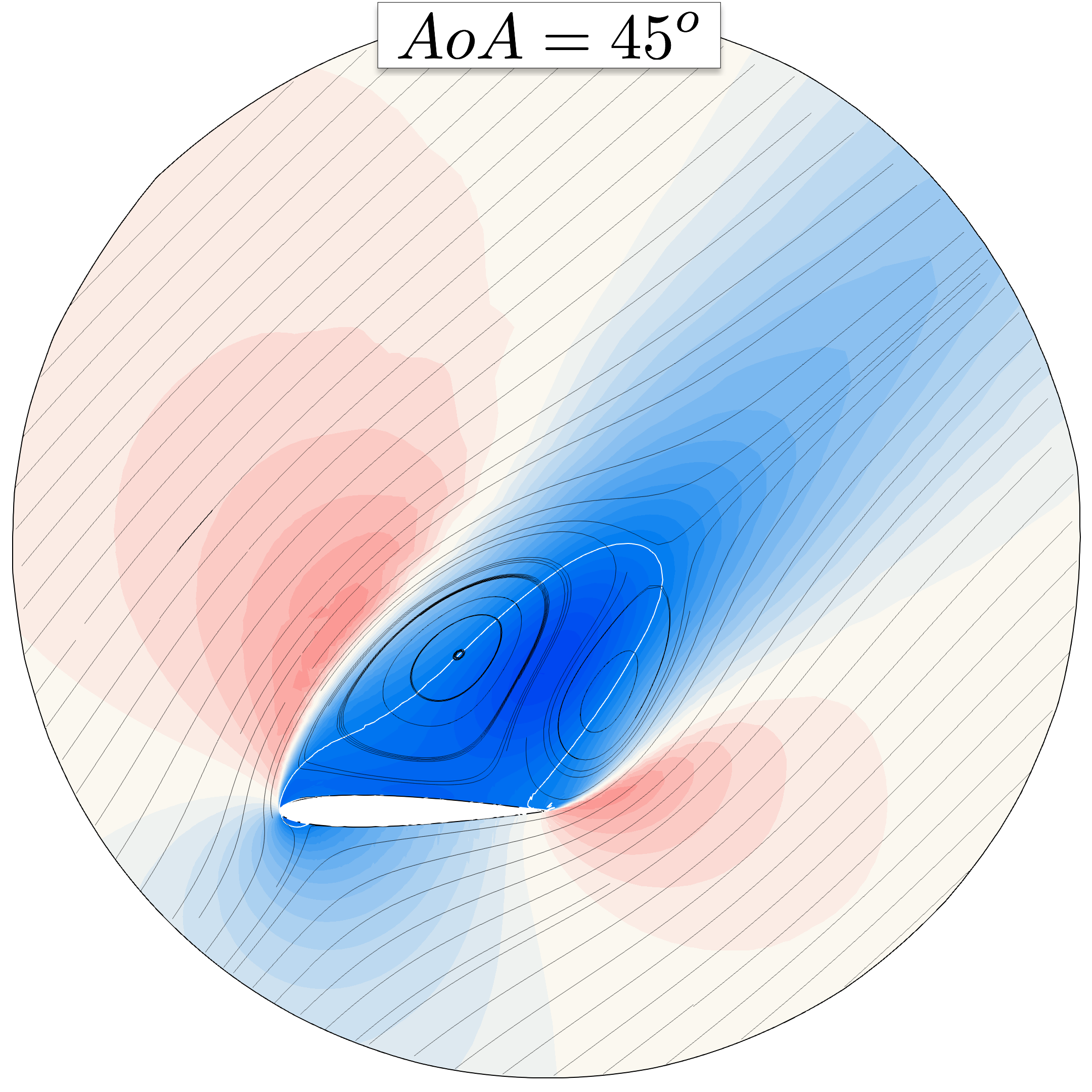}%
  \includegraphics[width=0.3\textwidth]{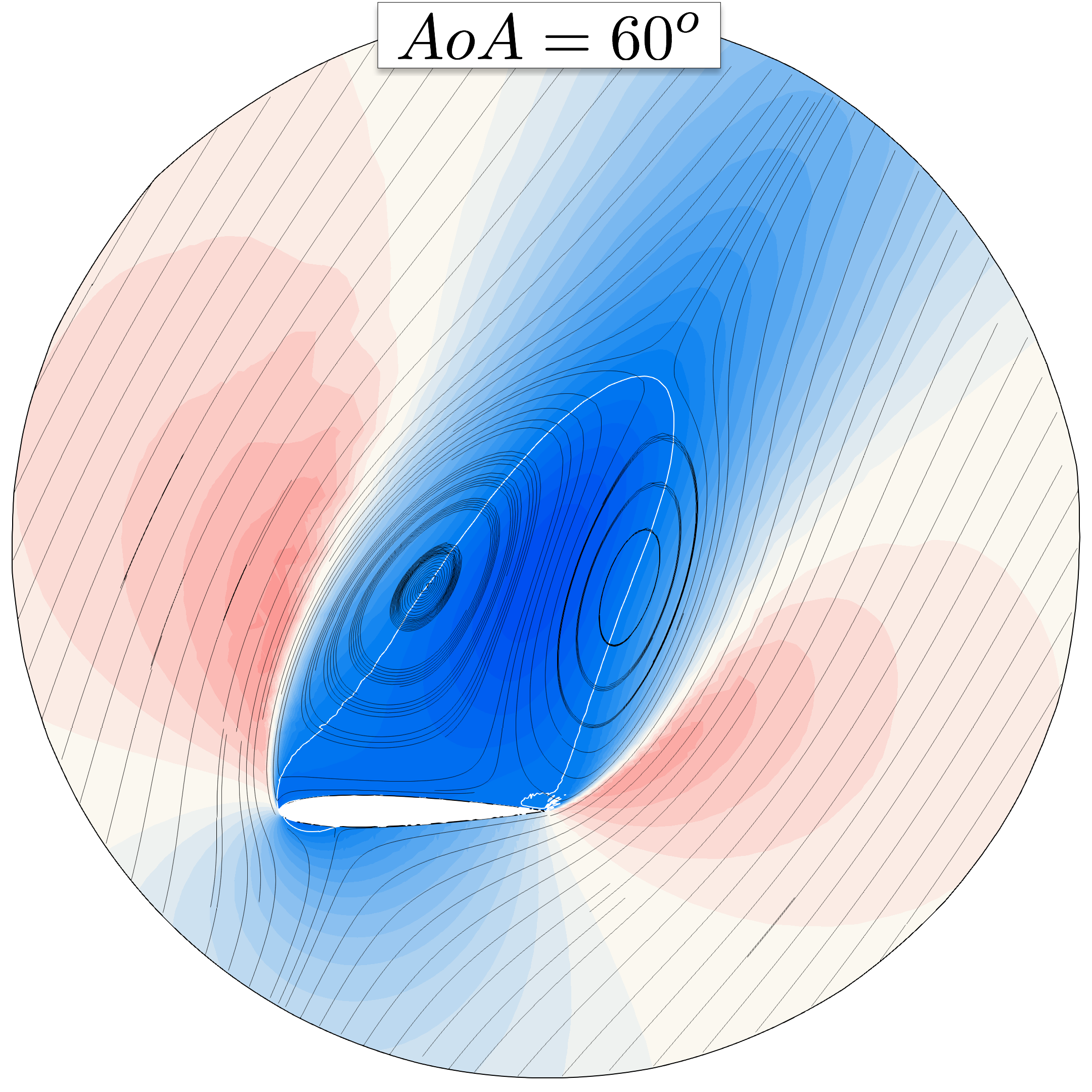}%
  \includegraphics[width=0.3\textwidth]{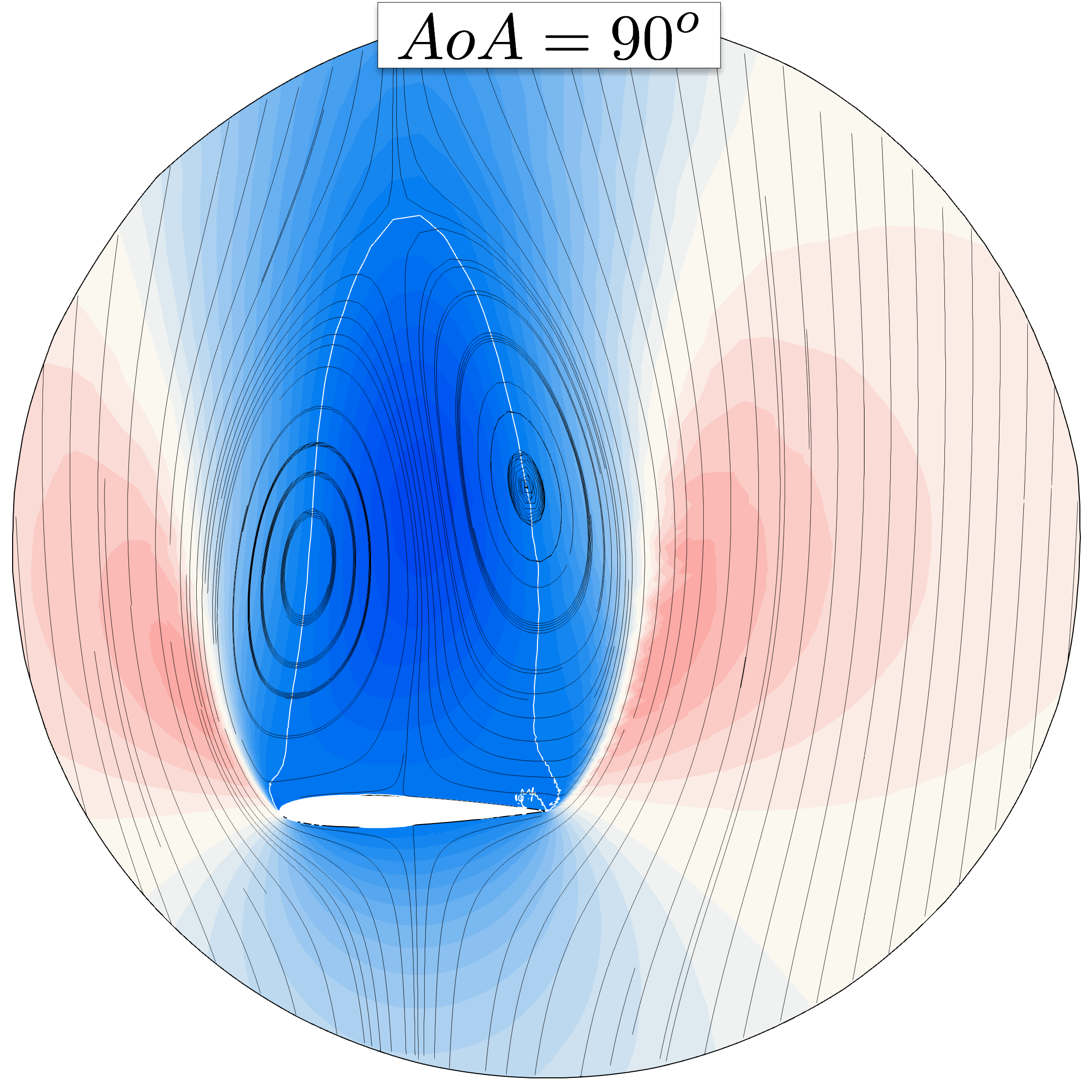}%
  \raisebox{.5\height}{\includegraphics[width=0.1\textwidth]{./figs/Wing_03_030}}
  \caption{Time-and-span averaged streamwise velocity $(\vec{U}\cdot
\hat{\theta}/U_o)$ for flow over \naca{0012} at $Re_c=2\times10^6$ using DDES;
black lines are the streamlines, and white contour lines of $\vec{U}\cdot
\hat{\theta}=0$ show separated region.}
\label{fig:naca_des_3d_avg}
\end{figure*}
%-----------------------------------------------------------------%
%-----------------------------------------------------------------%
\begin{figure*}[!t]
  \centering
  \includegraphics[width=1\textwidth]{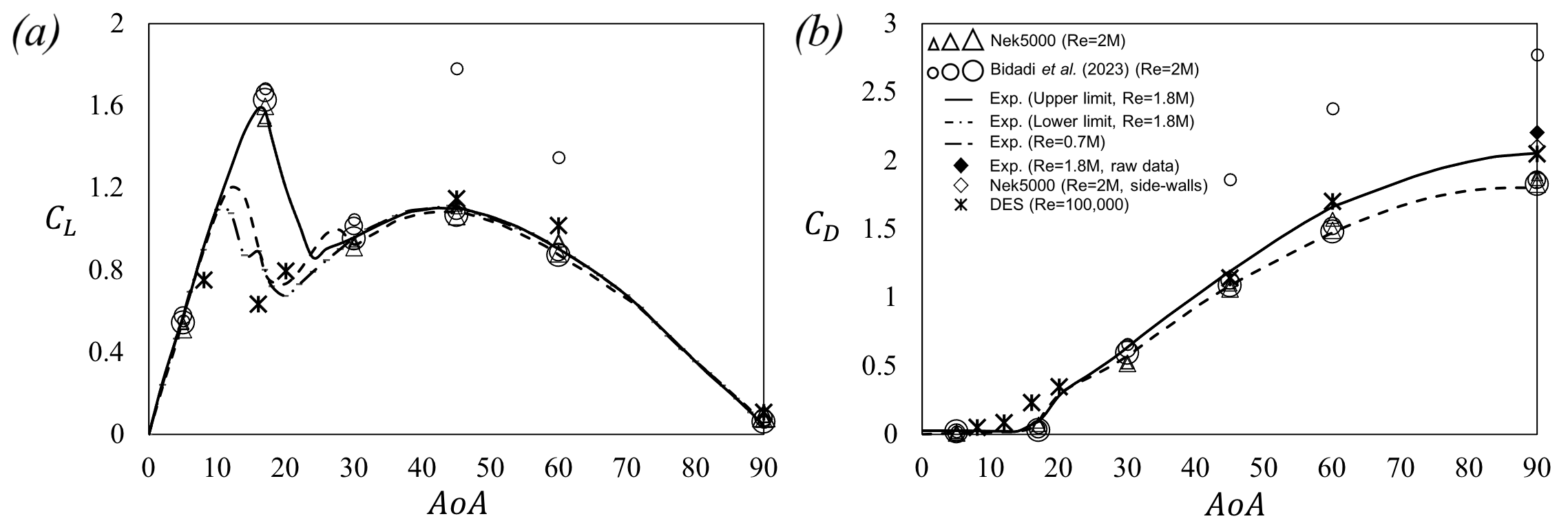}
  %\raisebox{1.1\height}{\includegraphics[width=0.2\textwidth]{./figs/Wing_04_01b}}
  \caption{\textit{(a)} Lift coefficient; \textit{(b)} drag coefficient as a
function of angle of attack for flow over \naca{0012} at $Re_c=2\times10^6$,
obtained with 3D-DDES runs; (open triangles) current data; (open circles) data
from \cite{Bidadi2023}; (solid line) upper limit of compiled experimental data;
(dashed line) lower limit of compiled experimental data; (dash-dotted line)
experimental data of \cite{Sheldahl1981}; (open diamond) ``CFD" tunnel results
at \aoa{90}; (crosses) DES data of \cite{Strelets2001}; (filled diamond) raw
data (without wind tunnel correction) from experiments of \cite{Critzos1955}.
\label{fig:cdcd_des_3d} }
  \label{fig:clcd_ddes_3d}
\end{figure*}
%-----------------------------------------------------------------%
% -----------------------------------------------------------------%
\subsubsection{Instantaneous flow \label{sec:edge_comp}}
%-----------------------------------------------------------------%
Figure~\ref{fig:qcrit_ddes_3d} shows the contours of the Q-criterion, 
defined as the second invariant of the velocity gradient tensor.  It is given by

\begin{equation}
  Q = 0.5\left( \lVert \bf{\Omega} \rVert^2 - \lVert \bf{S} \rVert^2 \right),
\end{equation}
where $\Omega$ is the rotation tensor and $S$ the strain tensor.  
We can observe the adequacy of the spanwise domain size for the present calculations.  
The structures for both the coarse and fine mesh are structurally similar although the fine mesh captures more small details of the flow.

Figure~\ref{fig:naca_des_3d} shows the instantaneous snapshots (z-slice) of the flow field 
for \aoa{45} case. Here, we show the projection of 
flow velocity vector on the unit vector in the direction of incoming flow,

$$  \hat{\theta} = \cos(AoA)\hat{i} + \sin(AoA)\hat{j}, $$
where $\hat{i},\hat{j}$ are the unit vectors in the x and y directions, respectively. 
We notice that the large scale vortex shedding at high angle affect the
distribution of the $k$ and $\tau$ significantly.  
Figure~\ref{fig:naca_tunnel_views} compares the instantaneous flow fields for
the unconfined and confined cases at \aoa{90}. We clearly observe that the
presence of side-walls accelerates the flow
(Fig.~\ref{fig:naca_tunnel_views} (a)) which is absent in the unconfined case 
(Fig.~\ref{fig:naca_tunnel_views} (b)). The blocking effect results in greater
coherence of structures in the wake region for the confined case. 
The model eddy viscosity, shown in
(Fig.~\ref{fig:naca_tunnel_views} (c,d)), shows this strengthening of vortical
strictres clearly, whereby magnitude and distribution is different between the
two case setups.   

% -----------------------------------------------------------------%
\subsubsection{Averaged flow and aerodynamic forces \label{sec:avg_flow}}
%-----------------------------------------------------------------%
Figure~\ref{fig:naca_des_3d_avg} shows the (time-and-span) averaged streamwise velocity 
for all the 3D cases.  Both the 5-degree and 17-degree cases remain fully attached.
As expected, we observe massive separated regions as the 
incoming flow angle increases.  For example, the separated region at $AoA=90^o$ extends for 
more than three chord lengths behind the airfoil section. 
  
Figure~\ref{fig:clcd_ddes_3d} presents a comparison of the averaged forces on
the airfoil surface. This includes our data (represented by triangles in
Fig.~\ref{fig:clcd_ddes_3d}) for the three grid
resolutions. We observe that the differences in results among the three
resolutions are minimal, even for higher angles, and we expect the
coarse resolution to be optimal for an accurate representation of the flow
physics. We also notice that while we observe a grid converged result at the
coarse level, \etal{Bidadi} (represented by circles in
Fig.~\ref{fig:clcd_ddes_3d}) obtained the converged 
results at the medium resolution. We believe the high-order convergence of the 
spectral element method is primarily responsible for this enhanced accuracy.
This has been shown to be the case for resolved LES by other studies comparing 
low-order codes with \nek{} \citep{Rezaeiravesh2021a}.

Overall, there is a good agreement between our converged values and those of
\etal{Bidadi}~(2023), for both lift and drag forces for all angles in
Fig.~\ref{fig:clcd_ddes_3d}. The region below the stall angle is sensitive to
the the flow setup. Hence, there is a significant difference between our data
and that of Strelets~\cite{Strelets2001} (represented by crosses in
Fig.~\ref{fig:clcd_ddes_3d}), who conducted the original DES at a magnitude
lower Reynolds number (\re{100,000}) than the present study. For higher angles,
all three numerical datasets yield similar behavior, although values by Strelets
is consistently overpredicted.  The main differece between the calculations is
the spanwise extent of the domain; Strelets performed calculation with spanwise
extent of only $1c$, which is heavily restricted when compared to our case and
that of \etal{Bidadi} (both studies use spanwise size of $4c$). We believe this
constriction of the domain to be the primarily responsible for the drag values
of Strelets to be consistently overpredicted.      

Figure~\ref{fig:clcd_ddes_3d} also compares our data with the experimental
datasets for equivalent Reynolds number and angle-of-attack configurations
available in the literature. However, the trend between the experiments and our
data is not so clear as it was with other numerical studies.  
%Therefore, it
%becomes instructive to take a closer look at the experimental setup and
%`corrections' the experiments employed.

The two lines (solid and dash-dotted) in Fig.~\ref{fig:clcd_ddes_3d} were
originally compiled by \cite{Strelets2001}.  They correspond to the experimental
results of \cite{abbott2012}; \cite{Critzos1955}; \cite{mccroskey1982}; and
\cite{mccroskey1987}, in which the averaged forces on \naca{0012} were
benchmarked at Reynolds numbers in the range \re{1--2\times 10^6}.  As pointed
out in \cite{Strelets2001}, there seems to be significant experimental scatter
for the lower angles ($AoA$ in the range $13^o--25^o$).  As discussed before,
this is to be expected since the flow is sensitive to the setup near the angle
of maximum lift.  Some of the factors responsible for this scatter include inlet
flow conditions, wing surface texture, and measurement apparatus.  A critical
assessment of these experimental datasets has been discussed extensively in
\cite{mccroskey1987}, where the authors point to the reliability of datasets of
\cite{abbott2012}, \cite{Critzos1955}, and \cite{mccroskey1982} in comparison
with other such datasets.  The post-stall results are taken primarily from the
study of \cite{Critzos1955} at \re{1.8\times 10^6}. 

According to expectation, our $C_l, C_d$ data for angles below the stall angle
matches the upper curve, which corresponds to experiments with flow tripping
(this study does not employ any laminar-to-turbulent transition model) and
low inflow turbulence levels.  For angles higher than the stall angle, on the
other hand, the corroboration in not good. With increasing angle of attack there
is significant underprediction of drag coefficient by the numerical simulations
at \re{2\times 10^6} when compared with the experiments. The maximum difference
of $\approx 11\%$ occours at the maximum angle, \aoa{90}.  The data for the
experiments shown in Fig.~\ref{fig:clcd_ddes_3d} has been corrected for the
blocking effect of the side walls.  The ``wall corrections" thus employed are
empirical at best and could be a possible source of differences observed. Since
\etal{Critzos}~\cite{Critzos1955} also shared their uncorrected data, we wanted
to replicate their wind tunnel experiments and assess the accuracy of such wall
corrections. 
We report the drag coefficient for the confined 3D case at \aoa{90} (open
diamond symbol) in Fig.~\ref{fig:clcd_ddes_3d}. It compares very well with
the raw data of \cite{Critzos1955} (filled diamond) with less than 3\%
difference.   
A comparision between the confined and unconfined cases as a
function of Reynolds number is performed in the next section.

%%%%-----------------------------------------------------------------%
%%%-----------------------------------------------------------------%
%%\begin{figure*}[!htb]
%%  \centering
%%  \includegraphics[width=1.0\textwidth]{./figs/WB_01a}
%%  \caption{Temporal evolution of drag coefficient for \textit{(a,b,c)}
%%Unconfined and \textit{(d,e,f)} confined wing for three Reynolds
%%numbers at \aoa{90}. }
%%\label{fig:block_fig1}
%%\end{figure*}
%%%-------------------------------%
%%
%%\begin{figure*}[!htb]
%%  \centering
%%  \includegraphics[width=1.0\textwidth]{./figs/WB_01b}
%%  \caption{Histogram (as percent occurrence) of drag coefficient for \textit{(a,b,c)}
%%Unconfined and \textit{(d,e,f)} confined wing for three Reynolds
%%numbers at \aoa{90}. }
%%\label{fig:block_fig2}
%%\end{figure*}
%%%-------------------------------%
%%
%%\begin{figure*}[!htb]
%%  \centering
%%  \includegraphics[width=1.0\textwidth]{./figs/WB_01c}
%%  \caption{Power spectral density of drag coefficient for \textit{(a,b,c)}
%%Unconfined and \textit{(d,e,f)} confined wing for three Reynolds
%%numbers at \aoa{90}. }
%%\label{fig:block_fig3}
%%\end{figure*}
%%%-----------------------------------------------------------------%
%-------------------------------%

\begin{figure*}[!htb]
  \centering
  \includegraphics[width=0.7\textwidth]{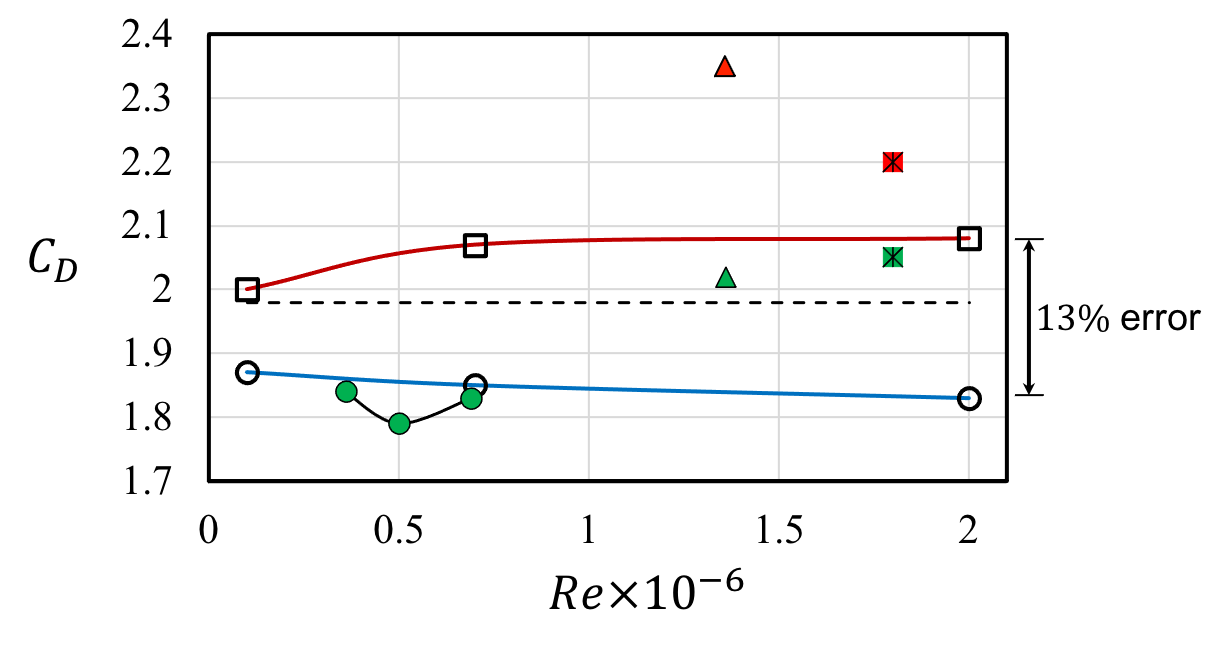}
  \caption{Drag coefficient as a function of Reynolds number at \aoa{90}; blue line:
unconfined wing; red line: confined wing; circles: exp. data of
\cite{Sheldahl1981}; triangles: exp. data of \cite{Critzos1955} obtained in
Langley 300 MPH-$7 \times 10$-foot tunnel; cross:  exp. data of
\cite{Critzos1955} obtained in Langley LPTP tunnel; green: corrected data; red:
raw data.}
\label{fig:block_fig4}
\end{figure*}
%-----------------------------------------------------------------%
%-----------------------------------------------------------------%

% -----------------------------------------------------------------%
\subsection{Assessment of confinement effect \label{sec:wall_block}}
%-----------------------------------------------------------------%

Figure~\ref{fig:block_fig4} shows the effect of confinement on the drag
coefficient for \naca{0012} at \aoa{90}. We tested three values of Reynolds
numbers from a low value of 100,000 to 2M.  We employed the same mesh for all
the simulations reported in Fig.~\ref{fig:block_fig4}. Since the resolution
was sufficient at \re{2M} it should also suffice the requirements at other
Reynolds numbers considered here.

We can observe different trends for
the drag on the confined vs the unconfined simulation. Whereas $C_d$ values for the
unconfined simulation (blue line) remains more or less constant (about $2\%$ variation) with
Reynolds number, the drag for the confined case (red line) shows slightly more sensitivity
($5\%$ increase with $Re$).  The differences between the two cases increases
slightly within the tested Reynolds number range, the maximum being around 13\%.     
As expected, the $C_d$ values are close to the drag on a 2D flat plate at $90^o$
incidence (dashed black line in Fig.~\ref{fig:block_fig4}) as studied by Hoerner \cite{hoerner1958}.
In Fig.~\ref{fig:block_fig4}, we also show data of some wind tunnel measurements. 
These include both uncorrected data (red symbols) and data corrected for wall
interference (green symbols). We observe a better match for the corrected values
of \cite{Sheldahl1981} with our open simulations at lower values of $Re$.  
At higher $Re$, raw $C_d$ values from the two experimental setups of 
\cite{Critzos1955} (filled triangles and crosses) are substantially
higher than our confined cases. The corresponding corrected values overpredict
our unconfined $C_d$ values by more than $7\%$. 

The difference in trends and relatively higher values of experiments of \cite{Critzos1955} 
point to the need for a through investigation, and possible reevaluation, of the
models of wall corrections employed in experiments. 
However, such a study needs data for different airfoils, at various angles of
attack and Reynolds number configurations.
It is outside the scope of the present investigation and will be taken up in the
future studies.    
        
\section{Conclusions \label{sec:conclusion}}

We developed a hybrid RANS-LES (HRLES) formalism that is particularly suited to
spectral element based CFD solvers. This formulation is based on the
recently proposed k-$\tau$ SST RANS model by
\etal{Tomboulides}~\cite{Tomboulides2024} which obviates the need of ad hoc
boundary conditions as required for k-$\omega$ class of models. The necessary
details are provided in Section~\ref{sec:methodology}.

We then presented a comprehensive assessment of 
the delayed detached eddy simulation (DDES) which uses the $k-\tau$ 
SST model for the near wall calculations.  We establish a formal strategy for NACA 
wing simulations, in general, by coupling the grid generation process for \nek{} 
with an open-source grid generator Construct2D \citep{construct2d}. A detailed 
discussion regarding the grid generation, mesh conversion, and mesh smoothing 
were presented in Section \ref{sec:grid_gen}. 

The verification of the $k-\tau$ SST RANS model was performed in Section~\ref{sec:rans_verify}. 
We compared our results with both the simulations of NASA's CFL3D code and
experiments and 
observed good agreement with CFL3D results. There were some discrepancies with
the experiments, which can be attributed to the limitations of the model, but in
general the corroboration is fair.

In the second part, we conducted an in-depth assessment of
sample requirements for statistical convergence and found that \nek{} requires
a fraction of the time needed by fully implicit low-order schemes to yield fully
converged results.  We also
demonstrated that the prediction of power spectral density with \nek{} is in
excellent agreement with experiments, as opposed to a representative
low-order code that overpredicted the power content (particularly
in the low frequencies) by an order of magnitude.
    
We later extended our study to the DDES of \naca{0012} at \re{2\times 10^6}
and for \aoa{0-90}. We demonstrated that the performance of \nek{} remains good 
for the hybrid calculations.  To assess the accuracy, we compared our results with 
those of Bidadi et al.~\cite{Bidadi2023}, who used a second-order finite-volume-based CFD 
code Nalu-Wind \citep{NaluWind} to conduct an Improved-DDES (IDDES) study of the same 
flow configuration.  From our thorough grid convergence study, we observe that 
\nek{} achieves converged results at one level coarser resolution than that of 
\etal{Bidadi} (2023). The converged results for both codes are in excellent 
agreement with each other.  

We also compared our DDES results with the experimental datasets available in the 
literature. For angles below the stall angle $\left( \alpha<20^o \right)$, we observed 
excellent agreement with datasets that were obtained with low-intensity turbulence inlet 
conditions and without flow tripping.  However, we observed significant differences between 
our unconfined results and the experimental wind tunnelresults at higher angles.  

We postulate that the majority of these differences come from the ``blocking effect" 
of the side walls in most of the experimental setups.  To check our hypothesis, 
we conducted separate calculations in a confined ``CFD" wind tunnel, with the aim of 
reproducing the findings of \cite{Critzos1955}. 
The comparison of our CFD-tunnel data with the raw data 
(uncorrected data for side walls) of \etal{Critzos} reveals much smaller differences than 
with unconfined configurations. 
Further analysis into confined simulations reveals drag at \aoa{90} to be a weak
function of the Reynolds number, whereby it increases marginally with $Re$.  
For unconfined simulations $C_d$ remained more or
less constant between $Re=10^5 -- 2\times 10^6$.  The largest difference between
the unconfined and confined cases is about 11\% and occurs at the maximum $Re$
considered in this study.  We also tested the uncorrected and corrected values for a few
experiments. They reveal some inherent inconsistencies, and the wall corrections
might need to be reevaluated taking into account the confined CFD simulations.
However, more work is required to address the issue in a systematic manner.
We note that since our primary aim was to include only  
the blocking effect of the side walls, we did not properly resolve the side walls to save 
on computational costs. This could be a source of some error between our results and those 
of \cite{Critzos1955}.    

We comment on some of the limitations of this study that could be improved in future work. 
For the O-grid mesh we employed a closed trailing-edge geometry.  
This is perhaps not the best strategy when it comes to the performance of the code. 
Using a curved trailing-edge permits \nek{} to use much higher simulation time steps 
and even higher CFL than 3.0 that the current study employed.  For the future, our target 
would be to use an order of magnitude higher CFL than the current study, by incorporating 
additional grid manipulation and smoothing tools in \nek{}.
For the confined simulations we used monodomain meshes. This is not the optimal
strategy as it requires meshing to be performed for every new flow angle. 
In the future, we propose to use an overlapping meshing strategy in \nek{} 
to perform calculations at several angles once a mesh has been generated.

\section*{Acknowledgments}
This material is based upon work supported by the U.S. Department of Energy,
Office of Science, under contract DE-AC02-06CH11357  and by the Exascale
Computing Project (17-SC-20-SC), a collaborative effort of two U.S.
Department of Energy organizations (Office of Science and the National
Nuclear Security Administration).
We gratefully acknowledge the computing resources provided on Improv (and/or Bebop and/or Swing),
a high-performance computing cluster operated by the Laboratory Computing Resource Center
at Argonne National Laboratory.
This research also used resources of the Argonne Leadership Computing Facility,
a U.S. Department of Energy (DOE) Office of Science user facility at Argonne National Laboratory
and is based on research supported by the U.S. DOE Office of Science-Advanced Scientific Computing
Research Program, under Contract No. DE-AC02-06CH11357.
We also acknowledge the help of Yu-Hsiang Lan who provided technical support during the simulations.
We gratefully acknowledge the help of Yuhsiang Lan who provided technical
support during the simulations.  

\bibliographystyle{unsrt}
\bibliography{bibs/emmd,bibs/ananias,bibs/abl,bibs/references.bib,bibs/vishal.bib}

\newpage
\appendix
\section{Numerical framework \label{sec:methodology_appn}}

Nek5000 is a spectral element method (SEM) code that is used for a wide range 
of thermal-fluids applications~\cite{nek5000}. 
Nek5000 has scaled to millions of MPI ranks using 
the Nek-based gsLib communication library to handle all near-neighbor and other 
stencil-type communications~\cite{fischer15}.
On CPUs, tensor contractions constitute the principal computational kernel 
(typically $>$ 90\% of the flops). These can be cast as small, dense matrix-matrix 
products resulting in high performance with a minor amount of tuning.

Nek5000 employs high-order spectral elements (SEs)~\cite{pat84} in which the solution, 
data, and test functions  are represented as {\it locally structured} $N$th-order 
tensor product polynomials on a set of $E$ {\it globally
unstructured} curvilinear hexahedral brick elements.   The approach yields two
principal benefits.  First, for smooth functions such as solutions to the
incompressible Navier-Stokes (NS) equations, high-order polynomial expansions
yield exponential convergence with approximation order, implying a significant
reduction in the number of unknowns ($n \approx EN^3$) required to reach
engineering tolerances.  Second, the locally structured forms permit local
lexicographical ordering with minimal indirect addressing and, crucially, the
use of tensor-product sum factorization to yield low $O(n)$ storage costs and
$O(nN)$ work complexities~\cite{sao80}.  As we demonstrate, the leading order
$O(nN)$ work terms can be cast as small, dense matrix-matrix products (tensor
contractions) with favorable $O(N)$ work-to-storage ratios (computational
intensity)~\citep{dfm02}.

%---------------------------------------------------------------------%
\subsection{Governing equations}
The incompressible NS equations are
\index{Navier-Stokes}
\begin{eqnarray}
    \label{eq:inc}
    \frac{\partial {u_j}}{\partial x_j} &=& 0, \\
    \label{eq:ns}
    \pp{u_i}{t} + u_j\frac{\partial {u_i}}{\partial x_j}
     &=& -\frac{1}{\rho}\frac{\partial {p}}{\partial x_i}
         %+2 \nu \frac{\partial S_{ij}}{\partial x_j} + f_i,
         + \nu \frac{\partial^2 u_i}{\partial x_j^2} + f_i,
\end{eqnarray}
where $u_i$ is the $i$th component of the velocity vector, $\rho$ is the
density, $p$ is the fluid pressure and $\nu$ is the fluid kinematic viscosity.
% and $S_{ij}= (\sfrac{1}{2}) \left( \sfrac{\partial{u_i}}{\partial
%x_j}+\sfrac{\partial{u_j}}{\partial x_i}\right)$ is the velocity strain tensor.

\subsection{Spatial discretization and temporal integration}

Time integration in Nek5000/RS is based on a semi-implicit splitting scheme
using $k$th-order backward differences (BDF$k$) to approximate the time
derivative coupled with implicit treatment of the viscous and pressure
terms and $k$th-order extrapolation (EXT$k$) for the remaining advection and forcing
terms.  This approach leads to independent elliptic subproblems comprising a
Poisson equation for the pressure, a coupled system of Helmholtz equations for
the three velocity components, and an additional Helmholtz equation for the
potential temperature.
The pressure Poisson equation is obtained by taking the
divergence of the momentum equation and forcing $\frac{\partial u_i^n}{\partial
x_i}=0$ at time $t^n=n \Delta t$. The velocity and temperature
Helmholtz equations are obtained once $p^n$ is known:
\begin{eqnarray} \label{pres}
-\frac{\partial^2}{\partial x_j\partial x_j} p^n\! &=&\! q^n, 
\\ \label{vel}
\frac{\beta_0}{\dt} u_i^n 
-\pp{}{x_j}\left(\frac{1}{Re}+\gamma \nu_t \right) 2S^n_{ij}
                              \! &=&\! - \pp{}{x_i} p^n + r_i^n, 
%\\ \label{temp}
%\hspace*{-.1in}
%\left[\frac{\beta_0}{\dt} -\pp{}{x_j}\left(\frac{1}{Pe}+ 
%   \frac{\gamma \nu_t}{Pr_t}\right) \pp{}{x_j}\right] \theta^n \! &=&\!  s^n,
\end{eqnarray}
where $\beta_0$ is an order-unity constant associated with BDF$k$
\citep{fischer17}, $S^n_{ij}$ is the resolved-scale strain-rate tensor as
described, and $q^n$, $r_i^n$ represent the sum
of the values from the previous timesteps for the contributions from BDF$k$ and
EXT$k$.  Also included in $r_i^n$ are eddy-diffusivity terms coming
from the mean-field eddy diffusivity mentioned above.  The fully coupled system
of Helmholtz equations for the three velocity components~(\ref{vel}) is used
only when the fluctuating (isotropic) part of the SGS stress tensor is modeled
using a Smagorinsky (SMG) model based on the fluctuating strain rate. When this
part is modeled through the use of a high-pass filter
(HPF)~\citep{Stolz_Schlatter_2005}, the resulting Helmholtz equations for the
three velocity components are not coupled.
%where $\beta_0$ is an order-unity constant associated with BDF$k$
%\citep{fischer17}, $S_{ij}$ is the resolved-scale strain-rate tensor as
%described, and $q^n$ and $r_i^n$, and $s^n$ represent the sum
%of the values from the previous timesteps for the contributions from BDF$k$ and
%EXT$k$.  Also included in $r_i^n$ and $s^n$ are eddy-diffusivity terms coming
%from the mean-field eddy-diffusivity mentioned above.  The fully coupled system
%of Helmholtz equations for the three velocity components~(\ref{vel}) is used
%only when the fluctuating (isotropic) part of the SGS stress tensor is modeled
%using a Smagorinsky (SMG) model based on the fluctuating strain rate. When this
%part is modeled through the use of a high-pass filter
%(HPF)~\citep{Stolz_Schlatter_2005} the resulting Helmholtz equations for the
%three velocity components are not coupled.

\noindent
With the given time splitting, we recast (\ref{pres})--(\ref{vel}) into weak
form and derive the spatial discretization by restricting the trial and test
spaces to be in the finite-dimensional space spanned by the spectral element
basis.   The discretization leads to a sequence of symmetric positive definite
linear systems for pressure and velocity.  The velocity system is diagonally
dominant and readily addressed with Jacobi-preconditioned conjugate
gradient iteration.  The pressure Poisson solve is treated with GMRES using
$p$-multigrid as a preconditioner.  Details of the formulation can be found in
\citep{fischer04,fischer17,malachi2022a}.

%---------------------------------------------------------------%
%---------------------------------------------------------------------%
\subsection{Large-eddy simulations}
Large eddy simulation (LES) employs a filtering technique to differentiate the
contributions of different scales. The filtering (of $\phi^*$) is represented
mathematically as 
\begin{equation}
\begin{aligned}
 & \hat{\phi}=\int_{-\Delta_f/2}^{\Delta_f/2}\; \mathcal{F}\left( x-\xi;x\right)\,\phi^*(\xi) \,d\xi, &
 \end{aligned}
 \label{eqn:Filtering}
\end{equation}
where $\Delta_f$ is the filter width. A spatially filtered resolved-scale formulation for the equations~(\ref{eq:inc})-(\ref{eq:ns}) can be written as
\begin{eqnarray}
\label{eq:filtered_inc}
    \frac{\partial {\hat u_j}}{\partial x_j} &=& 0, \\
    \label{eq:filtered_ns}
    \pp{\hat u_i}{t} + \hat u_j\frac{\partial {\hat u_i}}{\partial x_j}
     &=& -\frac{1}{\hat \rho}\frac{\partial {\hat p}}{\partial x_i}
        %+ 2\nu \hat S_{ij}
        + \nu \frac{\partial^2 u_i}{\partial x_j^2}
         -\pp{\hat \tau_{ij}}{x_j} + f_i,
\end{eqnarray}
where the extra stress term is given by
\begin{eqnarray}
         \hat \tau_{ij} = \widehat{u_iu_j} - \hat{u_i} \hat{u_j}.
\end{eqnarray}
$\hat \tau_{ij}$ is referred to as the {\it subgrid scale stresses} (SGS) 
and needs to be modeled from known flow quantities.

%---------------------------------------------------------------%
\subsection{High-Pass Filter model}

An alternative to the SGS model is the relaxation term (RT) model of
\etal{Stolz}~\cite{Stolz2005a}. The SGS stress terms
of~(\ref{eq:filtered_inc})-(\ref{eq:filtered_ns}) are replaced with a damping
factor acting only on the high wavenumber components of the solution, to arrive
at the high-pass-filter (HPF) LES model:
\begin{eqnarray}
\label{eq:hpf_inc}
    \frac{\partial {u_j}}{\partial x_j} &=& 0, \\
    \label{eq:hpf_ns}
    \pp{u_i}{t} + u_j\frac{\partial {u_i}}{\partial x_j}
     &=& -\frac{1}{\rho}\frac{\partial {p}}{\partial x_i}
         % + 2 \nu S_{ij} -\chi H_{N} * {u}_{i} + f_i,
         + \nu \frac{\partial^2 u_i}{\partial x_j^2} + f_i,
\end{eqnarray}
where $H_N$ is a high-pass filter (HPF) that is applied element-by-element and
$\chi$ is used as a weighting parameter.

The solutions in the SEM are locally represented on each element, $\Omega^e$,
$e=1,\dots,E$, as $N$th-order tensor-product Lagrange interpolation polynomials
on the Gauss-Lobatto-Legendre (GLL) nodes in the reference element, $\Oh := [-1,1]^3$.  
In fact, for low pass filtering, this representation can readily be expressed
as a tensor-product of Legendre polynomials, $P_k$.  In 1D, we can express
$u(x)=\sum_{k=0}^N {\tilde u}_k P_k(x)$ for the $N$th-order polynomial
approximation on $[-1,1]$. 
To construct the HPF, we start with a low-pass filter that dampens
the top $m$ modes quadratically by defining
 \begin{eqnarray}
  \hat u(x)= \mathcal{F}_N u(x) =\sum_{k=0}^N \sigma_k {\tilde u_k} P_k(x),
 \end{eqnarray}
\noindent
with $\{\sigma_k=\left(\frac{m-i}{m}\right)^2\}_{i=1}^m \mbox{ for } k=N-m+i$ and
$\sigma_k=1$ for $k \leq N-m$.
For example, we have $\sigma_6=(2/3)^2$, $\sigma_7=(1/3)^2$, and $\sigma_8=0$ when
using three modes ($m=3$) for $N=8$ and
$\sigma_7=(1/2)^2$ and $\sigma_8=0$ when using two modes ($m=2$) for $N=8$.
The low-pass filtered velocity then is used to construct the high-pass filter,
which is applied through the right-hand side in~(\ref{eq:hpf_ns}) as an additional term.  
If $F$ is the matrix operation that applies this one-dimensional low-pass filter,
then the three-dimensional low-pass filter $\mathcal{F}_N$ on $\Oh$ is given by
the Kronecker product, $F\otimes F \otimes F$, and the high-pass 
filter~\citep{Stolz_Schlatter_2005,Schlatter_Stolz_2006} is
\begin{eqnarray}
 %H_N * \bar{u} = \bar{u} - G * \bar{u} = \bar{u} - (F\otimes F \otimes F) \bar{u} \,,
 \mathcal{H}_N * u = u - \mathcal{F}_N u = u - (F\otimes F \otimes F)u.
\end{eqnarray}
This approach ensures stability, mitigates overshoot, and holds the
divergence-free condition \citep{Fischer2001,Negi2017a}.

For turbulence simulations,
the relaxation term $-\chi H_{N} * {u}_{i}$ provides the necessary drain of
energy out of a coarsely discretized system and, in the case of constant $\chi$,
can be interpreted as a secondary filter operation with the filter $Q_N * G$
applied every $1/(\chi \Delta t)$ time steps~\cite{Schlatter_Stolz_2004} with
$\Delta t$ being the stepsize of the time integration.  Here we use
$\chi=5$, which drains adequate energy to stabilize the simulations;
however, doubling the relaxation parameter to $\chi=10$ does not significantly
affect the results.  The high-pass filter used in polynomial space only affects
the last mode in the Legendre spectrum of the velocity or temperature (i.e., we take
$m=0$).

%\noindent
The high-wavenumber relaxation of the HPF model is similar to the
approximate deconvolution approach (ADM)~\cite{stolz2001}.  It is
attractive in that it can be tailored to directly act on marginally
resolved modes at the grid scale.
Following~\citep{Stolz_Schlatter_2005,Schlatter_Stolz_2006}, the approach
allows good prediction of transitional and turbulent flows with minimal
sensitivity to model coefficients.  Furthermore, the high-pass filters enable
the computation of the structure function in the filtered or HPF
structure-function model in all spatial directions even for inhomogeneous
flows, removing the arbitrariness of special treatment of selected (e.g.
wall-normal) directions.

\subsection{Standard $k-\omega$ model and its SST formulation}

The original $k-\omega$ model by Wilcox~\cite{Wilcox1998} solves two scalar
transport equations given by 
\begin{eqnarray}
    \frac{\partial \left(\rho k \right)}{\partial t} + \nabla \cdot (\rho k \vect{u}) &=& \nabla \cdot \Gamma_k \nabla k  + P_k - \rho \beta^* k \omega, \label{eqn:wilcox_k} \\
    \frac{\partial \left(\rho \omega \right)}{\partial t} + \nabla \cdot (\rho \omega \vect{u}) &=& \nabla \cdot \Gamma_\omega \nabla \omega + \alpha \frac{\omega}{k}P_k - \rho \beta \omega^2.
    \label{eqn:wilcox_w}
    \end{eqnarray}
    where $k=(1/2)(\langle u' \rangle + \langle v' \rangle + \langle w' \rangle)$ is the turbulent kinetic energy (TKE), $\omega$ is the rate of dissipation of TKE, and $P_k$ is the production of TKE. Here,
\begin{eqnarray}
  \Gamma_k &=&  \mu + \frac{\mu_t}{\sigma_k}, \nonumber \\
  \Gamma_\omega &=& \mu + \frac{\mu_t}{\sigma_{\omega}}.  \nonumber
\end{eqnarray}
In~(\ref{eqn:wilcox_k}--\ref{eqn:wilcox_w}), the production term is calculated as
$$ P_k = \min \left( \mu_t S^2, 10 C_\mu \rho k \omega \right),$$
where S is the magnitude of the strain-rate tensor.

In order to arrive at the SST model \citep{Menter1994}, a blending function is introduced in the $\omega-$equation as
\begin{eqnarray}
% \frac{\partial(\rho k)}{\partial t}+\nabla \cdot(\rho k  \vect{u}) &=& \nabla \cdot \Gamma_k \nabla k+P_k-\rho \beta^* k \omega \\
%
\frac{\partial(\rho \omega)}{\partial t}+\nabla \cdot(\rho \omega \vect{u}) &=& \nabla \cdot \Gamma_\omega \nabla \omega +\alpha \frac{\rho}{\mu_t} P_k-\rho \beta \omega^2 + 2(1-F_1) \rho \sigma_{\omega2} \frac{\nabla k \cdot \nabla \omega}{\omega}, 
\end{eqnarray}
and the eddy viscosity, closure coefficents, and auxiliary relations are defined as
\begin{eqnarray}
 \mu_t &=& \rho \frac{a_1 k}{\max \left(a_1 \omega, F_2 S\right)}, \\
 F_1 &=& \tanh \left(\operatorname{arg}_1^4\right), \\
 \arg_1 &=& \min \left[\max \left(\frac{\sqrt{k}}{\beta^* d \omega}, \frac{500 \nu}{d^2 \omega}\right), \frac{4 \sigma_{\omega2} k}{CD_{k \omega} d^2}\right], \\
C D_{k \omega} &=& \max \left(2 \sigma_{\omega2} \frac{\nabla k \cdot \nabla \omega}{\omega}, 10^{-10}\right), \\
F_2 &=& \tanh \left(\operatorname{arg}_2^2\right), \\
\operatorname{arg}_2 &=& \max \left(2 \frac{\sqrt{k}}{\beta^* d \omega}, \frac{500 \nu}{d^2 \omega}\right).
\end{eqnarray}
Here $d$ is the distance to the nearest wall.  The following model constants are used: 
\begin{eqnarray}
  \alpha_1 =5/9, \beta_1 =0.075, \sigma_{k1} =0.85, \sigma_{\omega1} =0.5, \nonumber \\
  \alpha_2 = 0.44, \beta_2 = 0.0828, \sigma_{k2} = 1, \sigma_{\omega2} = 0.856. \label{eqn:komega_ddes_const1}
\end{eqnarray}
Model constants are computed by a blend from the corresponding constants of the $k-\epsilon$ and $k-\omega$ model via $\alpha=\alpha_1 \cdot F1 + \alpha_2 \cdot (1-F1)$,  and so on.

\subsection{$k-\omega$ SST DDES formulation}
%----------------------------------------------------------%

The governing equation for $k$ for the SST-DDES formulation~\citep{Menter2003} is defined by
\begin{eqnarray}
  \frac{\partial \left(\rho k \right)}{\partial t} + \nabla \cdot (\rho k \vect{u}) &=& \nabla \cdot \Gamma_k \nabla k + P_k - \rho k^{3/2}/l_{DDES}, \label{eqn:komega_des_k}
    %
%    \frac{\partial \left(\rho \omega \right)}{\partial t} + \nabla \cdot (\rho \omega \vect{u}) &=& - \nabla \cdot \left[ \left( \mu + \sigma_\omega \mu_t \right) \nabla \omega \right] + \alpha \frac{\rho}{\mu_t} P_k - \beta \rho \omega^2 \nonumber \\
%     &+& 2(1-F1)\rho \sigma_{\omega 2} \frac{\nabla k \cdot \nabla \omega}{\omega} \label{eqn:komega_des_omega} \\
    %
%    \mu_t &=& \rho \frac{a_1 \cdot k}{\max \left( a_1 \cdot \omega, F2 \cdot S\right)} 
    \label{eqn:komega_ddes}
\end{eqnarray}
%where $F_1$ and $F_2$ are the blending functions given by,
%
%\begin{eqnarray}
%  F_1   &=& \tanh \left(arg_1^4\right)\\
%  arg_1 &=& \min \left( \max \left( \frac{\sqrt{k}}{C_\mu d}, \frac{500\nu}{d^2 \omega}\right), \frac{4\rho\sigma_{\omega 2} k}{CD_{k\omega} d^2} \right) \\
%  CD_{k\omega} &=& \max \left( 2\rho \sigma_{\omega 2} \frac{\nabla k \cdot \nabla \omega}{\omega}, 10^{-10} \right) \\
%  F_2   &=& \tanh \left( arg_2^2 \right) \\
%  arg_2 &=& \max \left( \frac{2\sqrt{k}}{C_\mu \omega d}, \frac{500 \nu}{d^2 \omega} \right)
%\end{eqnarray}
%
%
where the DDES length scale $(l_{DDES})$ is given as,

\begin{eqnarray}
  l_{DDES} &=& l_{RANS} - f_d \max \left(0, l_{RANS} - l_{LES} \right), \\
  l_{LES} &=& C_{DES} h_{max}, \\
  l_{RANS} &=& \frac{\sqrt{k}}{C_\mu \omega}, \\
  C_{DES} &=& C_{DES1} \cdot F_1 + C_{DES2} \cdot \left( 1-F_1 \right).
\end{eqnarray}
Here $h_{max}$ is the maximum edge length of the cell, and $f_d$ is the empirical blending function that is computed, following \citep{Strelets2001}, as
\begin{eqnarray}
  f_d &=& 1-\tanh \left[ \left( C_{d1} r_d\right) ^ {C_{d2}}\right], \\
  r_d &=& \frac{\nu_t + \nu}{\kappa^2 d^2 \sqrt{0.5 \left( S^2 + \Omega^2 \right)}}  .
\end{eqnarray}
Here, $S$ is the magnitude of the strain-rate tensor, and $\Omega$ is the magnitude of the vorticity tensor. 
The model constants read as follows:
\begin{eqnarray}
  C_\mu = 0.09, \; \kappa = 0.41, \; a_1 = 0.31, \label{eqn:komega_ddes_const0}\\
  C_{DES1} = 0.78, \; C_{DES2} = 0.61, \; C_{d1} = 20, \; C_{d2} = 3. 
\end{eqnarray}

\end{document}